\documentclass[%
 reprint,
nofootinbib,
 amsmath,amssymb,
 aps,
 prd,
]{revtex4-2}

\usepackage{graphicx}
\graphicspath{{figures/}{./}}
\usepackage{dcolumn}
\usepackage{bm}
\usepackage[colorlinks = true,
            linkcolor = black,
            urlcolor  = blue,
            citecolor = blue,
            anchorcolor = black]{hyperref}
\usepackage[mathlines]{lineno}
\usepackage{aas_macros}
\usepackage{xspace}
\usepackage{multirow}
\usepackage{tikz,xcolor}

\newcommand{\mosfit}{\textsc{MOSFiT}\xspace}
\newcommand{\gr}{$\gamma$-ray\xspace}
\newcommand{\Grs}{$\gamma$~rays\xspace}
\newcommand{\fermi}{\textit{Fermi}\xspace}
\newcommand{\fermiLAT}{\textit{Fermi}-LAT\xspace}
\newcommand{\FermiLAT}{\textit{Fermi}~LAT\xspace}
\newcommand{\probtot}{88.5\,\%}
\newcommand{\tsmax}{12.4}
\newcommand{\tsmaxsigma}{3.5}

\newif\ifArxiv
\Arxivtrue

\ifArxiv
\definecolor{lime}{HTML}{A6CE39}
\DeclareRobustCommand{\orcidicon}{
	\begin{tikzpicture}
	\draw[lime, fill=lime] (0,0) 
	circle [radius=0.16] 
	node[white] {{\fontfamily{qag}\selectfont \tiny ID}};
	\draw[white, fill=white] (-0.0625,0.095) 
	circle [radius=0.007];
	\end{tikzpicture}
	\hspace{-2mm}
}
\foreach \x in {A, ..., Z}{\expandafter\xdef\csname orcid\x\endcsname{\noexpand\href{https://orcid.org/\csname orcidauthor\x\endcsname}
			{\noexpand\orcidicon}}
}
\newcommand{\orcidauthorA}{0000-0002-0738-7581} 
\newcommand{\orcidauthorB}{0000-0003-4743-1679} 
\newcommand{\figsens}{15}

\else

\newcommand{\figsens}{\ref{fig:sensitivity}}

\fi

\begin{document}

\title{Search for Axionlike-Particle-Induced Prompt $\gamma$-Ray Emission from Extragalactic Core-Collapse Supernovae with the \emph{Fermi} Large Area Telescope}

\author{Manuel Meyer
\ifArxiv
\orcidA{}
\fi
}
 \email{manuel.e.meyer@fau.de}
\affiliation{%
Erlangen Centre for Astroparticle Physics, University of Erlangen-Nuremberg, Erwin-Rommel-Str. 1, 91058 Erlangen, Germany
}
\affiliation{
W. W. Hansen Experimental Physics Laboratory, Kavli Institute for
Particle Astrophysics and Cosmology, Department of Physics and SLAC National Accelerator Laboratory, Stanford University, Stanford, California
94305, USA
}%

\collaboration{\emph{Fermi}-LAT Collaboration}

\author{Tanja Petrushevska
\ifArxiv
\orcidB{}
\fi
}
\email{tanja.petrushevska@ung.si}
\affiliation{
 Centre for Astrophysics and Cosmology, University of Nova Gorica, Vipavska 11c, 5270 Ajdov\v{s}\u{c}ina, Slovenia
}%

\date{\today}

\begin{abstract}
During a core-collapse supernova (SN), axionlike particles (ALPs) could be produced through the Primakoff process and subsequently convert into \Grs in the magnetic field of the Milky Way. 
We do not find evidence for such a \gr burst in observations of extragalactic SNe with the \fermi Large Area Telescope (LAT).
The SN explosion times are estimated from optical light curves 
and we find a probability of about $\sim 90\,\%$ that the LAT observed at least one SN at the time of the core collapse.
Under the assumption that at least one SN was contained within the LAT field of view, 
we exclude photon-ALP couplings $\gtrsim 2.6\times10^{-11}\,\mathrm{GeV}^{-1}$ for ALP masses $m_a \lesssim 3\times 10^{-10}$\,eV, within a factor of $\sim 5$ of previous limits from SN1987A. 
\end{abstract}

\maketitle


\section{Introduction}
\label{sec:intro}

Axionlike particles (ALPs) are hypothetical pseudo Nambu Goldstone bosons predicted in numerous extensions of the standard model~\cite{jaeckel2010} and are particle candidates for dark matter~\cite{preskill1983,dine1983,abbott1983,arias2012}.
They are closely related to the axion, which was originally proposed to solve the strong \textit{CP} problem~\cite{pq1977,weinberg1978,wilczek1978}. 
Axions and ALPs could be detected through the coupling to photons
given by the Lagrangian density $\mathcal{L} = g_{a\gamma} \mathbf{E}\cdot\mathbf{B}a$, where 
$g_{a\gamma}$ is the coupling strength, $\mathbf{E}$ is the electric field of the photon, $\mathbf{B}$ is an external magnetic field, and $a$ is the ALP field strength~\cite{raffelt1988}. 

For low mass ALPs, $m_a \lesssim 10^{-9}$\,eV, stringent constraints on $g_{a\gamma}$ were derived from the nonobservation of a \gr burst from the core-collapse supernova (SN) SN1987A, which occurred at a distance of $\sim50\,$kpc~\cite{2015JCAP...02..006P}. 
Such ALPs could be produced through the conversion of 
thermal \Grs, created during the core collapse, into ALPs in the electrostatic fields of protons and ions, i.e., the Primakoff process.
The resulting ALP rate, $d\dot{N}_a/dE$, follows a thermal spectrum and peaks at $\sim60$\,MeV for a progenitor mass of $10\,M_\odot$ MeV~\cite{2015JCAP...02..006P}.
These ALPs would escape the core, and could convert back into photons in astrophysical magnetic fields with a conversion probability $P_{a\gamma}$. 
The flux of the resulting prompt \gr emission can be written as~(e.g., \cite{2017PhRvL.118a1103M})
\begin{equation}
    \frac{d\phi}{dE} = \left(\frac{g_{a\gamma}}{g_0}\right)^4\frac{P_{a\gamma}(g_0,m_a)}{4\pi d^2} \frac{d\dot{N}_a}{dE}(g_0),
    \label{eq:alpspec}
\end{equation}
where $d$ is the luminosity distance to the SN. 
The \gr spectrum depends on  $g_{a\gamma}^4$, with two powers of $g_{a\gamma}$ coming from the ALP production and two powers from the photon-ALP conversion (evaluated at some reference coupling $g_0$ in the equation above).

The \Grs should arrive simultaneously with the neutrinos produced in the SN, which therefore provide a time stamp to search for the \gr emission.
In contrast, ordinary \gr bursts associated with SNe
should be delayed relative to the core collapse 
by at least tens of seconds or even hours,
since the jet needs to reach the surface of the collapsing star first (see, e.g., \cite{Kistler:2012as,woosley_2012}).

Furthermore, the spectra of ordinary bursts peak at tens or hundreds of keV and have a very different shape compared to the ALP-induced one~\cite{hjorth_bloom_2012}.
Therefore, a \gr burst detected above tens of MeV in co-incidence with the neutrinos would be a ``smoking gun'' signature for ALPs.

In the case of a Galactic SN, 
the \fermi Large Area Telescope (LAT), which detects \Grs between 20\,MeV and beyond 300\,GeV~\cite{2009ApJ...697.1071A}, is expected to be more than an order of magnitude more sensitive to the prompt \gr emission compared to the SN1987A constraints~\cite{2017PhRvL.118a1103M}.
However, given the fact that SNe should occur at a rate of
$\sim3$ per century
in the Milky Way (see, e.g.,~\cite[][]{adams2013}) and that the field of view of the LAT is of the order of $\sim20\,\%$ of the sky, a detection within the lifetime of \fermi seems unlikely. 

In this Letter, we search for the prompt \gr emission of known extragalactic SNe that have occurred during the \fermi mission. 
As current generation neutrino telescopes are not sensitive enough to detect events from such sources~(e.g., \cite{kistler2011}), we estimate the explosion time from light curves obtained with optical transient facilities.
For well-sampled light curves it has been shown that simple analytic functions can provide estimates for the onset time of the optical emission within an accuracy of less than a day~\cite{cowen2010}. 

The Letter is structured as follows. In  Sec.~\ref{sec:snsample}, we describe the SN sample selection and our optical light curve fitting procedure. 
The \gr analysis and its results are presented in Sec.~\ref{sec:gray}.
We discuss and summarize our findings in Sec.~\ref{sec:concl}.
Further details on the analysis of optical light curves and the \gr emission are provided in the Supplemental Material (SM), which includes the additional references \cite{2011ApJ...728...14P,2014ApJS..213...19B,2011ApJ...728...14P,2014ApJS..213...19B,2011MNRAS.413.2583S,2011MNRAS.416.3138V,2011ApJ...740...41C,2012ApJ...755..161K,2012ApJ...749L..28V,2016AA...593A..68F,2015ApJ...799...51M,2012ApJ...760L..33B,2016AA...593A..68F,2016ApJ...821...57D,2018ApJ...863..109Z,2018MNRAS.475.2591S,2014ApSS.354...89B,2016AA...592A..89T,2018MNRAS.478.4162P,2018ApJ...860...90V,1982ApJ...253..785A,2008MNRAS.383.1485V, 2015A&A...574A..60T,Higson2019,2011ATel.3288....1G,2012ATel.3881....1A,2013ATel.5137....1C,2017ATel10481....1I,2019MNRAS.485.1559P,40mprogenitor,2019ApJS..243....6G,2017arXiv170804058L,2019arXiv190402171F}.

\section{Supernova sample and determination of explosion time from optical light curves}
\label{sec:snsample}

We select SNe that are publicly available in the Open Supernova Catalog (OSC)~\cite{
2017ApJ...835...64G}\footnote{\url{https://sne.space/}} that (a) have been detected after the start of the \fermiLAT science operations (08/03/2008);
(b) are core-collapse SNe of type Ib or Ic. The progenitors for these types of SNe are Wolf-Rayet stars or blue supergiants for which the delay between core collapse and the the onset of the optical emission should be of the order of hours or less~\cite{Kistler:2012as}; (c) are located at a redshift $z < 0.02$. For the corresponding luminosity distance of $d \lesssim 80\,$Mpc constraints on $g_{a\gamma}$ better than previous limits, $g_{a\gamma} < 6.6\times10^{-11}\,\mathrm{GeV}^{-1}$~    \cite{2017NatPh..13..584A}, should still be possible, since the \gr flux scales as $g_{a\gamma}^4 / d^2$ and limits of the order of $g_{a\gamma} \lesssim 2\times10^{-12}\,\mathrm{GeV}^{-1}$ should be possible for an SN in M31 ($d_\mathrm{L} = 778\,$kpc)~\cite{2017PhRvL.118a1103M}.

We inspect the optical light curves from this sample taken from the OSC, and select those SNe that have a sufficient number of premaximum data points that enable us to determine the explosion time. 
We \textit{a posteriori} add two SNe (SN2010bh, PTF15dtg) that occured at larger distances (but still with $z\lesssim0.06$ or $d \lesssim 270\,$Mpc) due to their well sampled light curves. 
This leaves us with 20 SNe that are listed in the SM.  
We also added additional pre-explosion flux upper limits from the literature, which are not listed in the OSC.

\begin{figure*}[tbh]
    \centering
    \includegraphics[width = .99\linewidth]{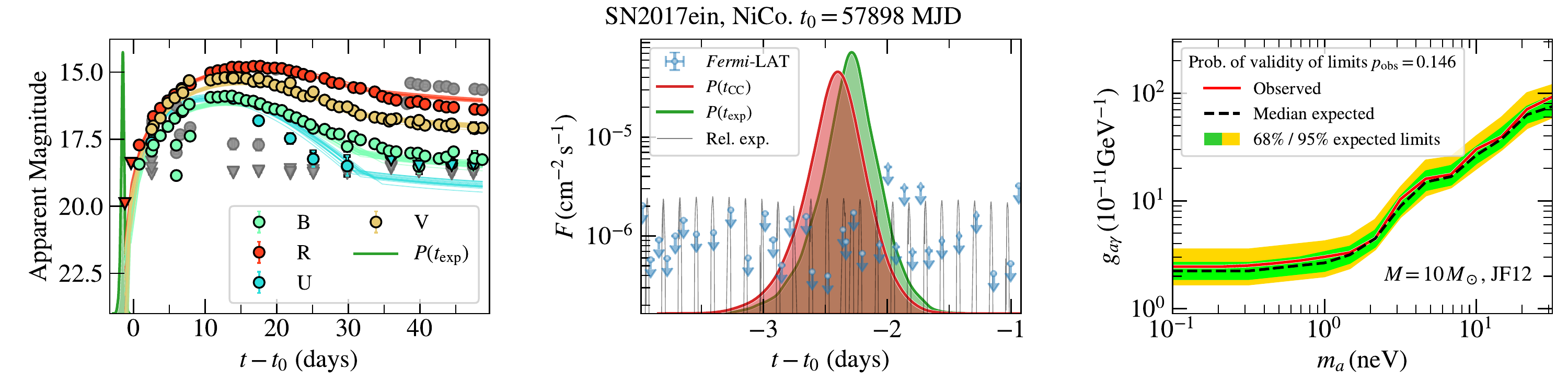}
    \caption{The three steps in the analysis procedure for one example, SN2017ein. \textit{Left:} the \mosfit optical light curve fit with the NiCo model. Measurements at different optical bands are shown as colored markers. 
    Grey markers indicate bands that are excluded from the fit since they do not contain sufficient data points pre- and post-maximum.
    The marginalized posterior for the SBO time, $P(t_\mathrm{exp})$, is shown as a green filled curve.
    \textit{Center:} the \gr light curve for the first energy bin between 60 and 83\,MeV together with the posteriors for the SBO and core collapse times (green and red curves, respectively).
    The normalized LAT exposure for each orbit is shown as gray thin lines. 
    \textit{Right:} Upper limits on the photon-ALP coupling derived from the non-observation of the \gr burst under the assumption that the SN was in the field of view of the LAT during the core collapse. The observed limit is shown in red, the expected limits are shown as colored bands, and the dashed black line.}
    \label{fig:lcs}
\end{figure*}

In order to determine the explosion time from the optical light curves, we use the  \mosfit code\footnote{\url{https://mosfit.readthedocs.io/}} (Modular Open Source Fitter for Transients, \citep{2018ApJS..236....6G}). \mosfit is an open-source code that adopts a Markov Chain Monte Carlo  approach to fit multiband light curves, and provides the posterior probability distributions for the free parameters in the model. 
We assume the radioactive nickel-cobalt decay model (NiCo) \citep{1994ApJS...92..527N,2017ApJ...849...70V}, which successfully describes the power source in normal Type~I SNe that lack hydrogen lines in their spectra. Unlike Type Ia SNe which are the thermonuclear explosions of white dwarfs, Type Ib/c SNe occur when stripped-envelope massive stars undergo core-collapse at the end of their lives. 
The bolometric luminosity in the model depends on the nickel mass synthesized in the explosion and the assumptions on the heating source can be found in \cite{2017ApJ...849...70V,2017ApJ...850...55N} and in the SM.
We also test a modified version of the NiCo model, where we add an additional component for the shock breakout (SBO) phase~\cite{2007ApJ...667..351W,cowen2010}, which we call SBO+NiCo, and a generic exponential rise and power-law decay model (exppow).
For each model, we extract the marginalized posterior for the explosion time $P(t_\mathrm{exp})$.
An example of the fit of the NiCo model to the optical light curve is shown in Fig.~\ref{fig:lcs} (left panel) for SN\,2017ein. 
In general, we find very good agreement with previously published values of $t_\mathrm{exp}$.

The SBO will be delayed with respect to the time of the core collapse, $t_\mathrm{CC}$, when we expect the ALP emission, due to the shock propagation through the stellar envelope~\cite{2013ApJ...778...81K}.
For this reason, we convolve $P(t_\mathrm{exp})$ with a 
boxcar function between 40\,s and $2\times10^4$\,s
to arrive at the posterior for $P(t_\mathrm{CC})$ (see the green and red lines in Fig.~\ref{fig:lcs}, center panel).
The time delay values are motivated from Ref.~\cite[see their  Fig.~2]{2013ApJ...778...81K} for Wolf-Rayet stars and blue supergiants.
The median values of $t_\mathrm{CC}$ for the convolved marginalized posterior are reported in the SM together with the time interval ($t_{\mathrm{CC},95}$) that makes up the 95\,\% quantile of $P(t_\mathrm{CC})$.

\section{Gamma-ray analysis}
\label{sec:gray}

Having established a time window for the likely arrival time of the ALP-induced \gr burst, we continue by extracting \gr events observed with the \FermiLAT between 60-600\,MeV belonging to the  \texttt{P8R3\_SOURCE\_V2} class with a zenith angle $\leqslant 80^\circ$ in a $20^\circ\times20^\circ$ region of interest (ROI) centered on the SN coordinates.
We extend the time interval to 60\,days around the date of the SN discovery in order to derive an average ROI model to describe the emission from the background sources and to extract the expected sensitivity (see below).
At 60\,MeV the contamination of diffuse background photons is already significant and the point spread function of the LAT is large (the 68\,\% containment angle is close to $10^\circ$)\footnote{See \url{https://www.slac.stanford.edu/exp/glast/groups/canda/lat_Performance.htm}} and we do not attempt an analysis at lower energies. 
We do not expect significant ALP emission above $\sim500\,$MeV for a $10\,M_\odot$ progenitor~\cite{2015JCAP...02..006P}.
For such a progenitor mass, the chosen energy range encompasses approximately $90\,\%$ of the energy released by the SN in form of ALPs.
In contrast, the energy range of the \gr burst monitor (GBM) on board the \fermi satellite only covers about $1\,\%$ of the released energy. 

We first derive an average model for each ROI using the \textsc{fermitools}\footnote{\url{https://fermi.gsfc.nasa.gov/ssc/data/analysis/}} v1.0.0 and \textsc{fermipy}\footnote{\url{https://fermipy.readthedocs.io/}} v.0.17.4~\cite{2017ICRC...35..824W} by considering all point sources within a $30^\circ\times30^\circ$ region around the SN, which are included in the \fermiLAT 8\,yrs source list,\footnote{\url{https://fermi.gsfc.nasa.gov/ssc/data/access/lat/fl8y/}} as well as templates for the Galactic and isotropic diffuse emission.\footnote{\url{https://fermi.gsfc.nasa.gov/ssc/data/access/lat/BackgroundModels.html}}
Spectral parameters of sources within $5^\circ$ of the ROI center are left free to vary as well as normalizations of sources up to $10^\circ$ and of the background templates. 
The SN spectrum in Eq.~\ref{eq:alpspec} depends on the progenitor mass through $d\dot{N}_a/dE$, and we conservatively assume a $10\,M_\odot$ progenitor for the entire analysis (rates for higher masses the rates for 10, 18, and 40\,$M_\odot$ progenitors are shown in the SM).
We conservatively include only the photon-ALP oscillations in the
coherent component of the
Galactic magnetic field (GMF) of the Milky Way, which is modeled as in Ref.~\cite[][hereafter JF12]{2012ApJ...761L..11J} and 
predicts magnetic fields of the order of $\mathcal{O}(\mu\mathrm{G})$.\footnote{Including the magnetic field in the host galaxy would increase the number of \Grs and thereby improve the LAT's sensitivity for a detection~\cite{2017PhRvL.118a1103M}.
We also neglect the turbulent component of the GMF, as its coherence length is too small to induce significant oscillations~\cite{2012PhRvD..86g5024H}. 
}
With this model, we evaluate $P_{a\gamma}$ following Refs.~\cite{2012PhRvD..86g5024H,2014JCAP...09..003M}. 
The spectral normalization, which is proportional to $g_{a\gamma}^4$, see Eq.~\eqref{eq:alpspec}, of this template is left free to vary.

After optimizing the average model,
we calculate the \gr light curve in eight logarithmic energy bins of the potential SN for each \textit{good time interval} (GTI), i.e., the time interval where the SN position was in the field of view of the LAT.  
We additionally extract the likelihood values as a function of source flux or equivalently $g_{a\gamma}$ for each SN ($i$), GTI ($j$), and energy bin ($k$), for which we use the shorthand notation $\mathcal{L}_{ijk}(g_{a\gamma})\equiv \mathcal{L}_{ik}(g_{a\gamma}, t_j, \boldsymbol{\theta}|D_k)$, where $t_j$ is the central time of the GTI, $\boldsymbol{\theta}$ are nuisance parameters, i.e., the spectral parameters of the other sources in the ROI, and $D_k$ is the data. 
Since the likelihoods are extracted in narrow bins of energy, they are independent of the assumed spectral shape~\cite{2015PhRvL.115w1301A}, which allows us to test different spectral models for the \gr burst without having to repeat the light curve extraction.
In the light curve extraction we assume $m_a \lesssim 1\,$neV (our final results take the full dependence of $P_{a\gamma}$ on $m_a$ into account).
From the equations of motion of the photon-ALP system one can derive that $P_{a\gamma}$ is constant up to a certain value of $m_a$ and starts to decrease for higher values.
This is the case when $m_a^2\gg g_{a\gamma} B E \sim 0.5\,\,$neV~\cite{raffelt1988},  
For $E=100$\,MeV, $B=1\,\mu$G, and $g_{a\gamma} = 10^{-11}\,\mathrm{geV}^{-1}$.
The light curve for the first energy bin is shown in Fig.~\ref{fig:lcs} (central panel) together with the normalized exposure for each GTI (grey lines), and the posterior functions extracted from \mosfit and smeared with the boxcar function.

The significance of the ALP signal is given by the test statistic $\mathrm{TS}_{ij} = -2 \sum_k(\ln\mathcal{L}_{ijk}(g_{a\gamma} = 0) - \ln\mathcal{L}_{ijk}(\hat{g}_{a\gamma}))$, where in the first term the likelihood has been maximized on the condition that $g_{a\gamma} = 0$ and in the second term the likelihood is unconditionally maximized resulting in the best-fit value $\hat{g}_{a\gamma}$ for the coupling.
The highest observed $\mathrm{TS}$ value is found for SN2011bm with $\mathrm{TS}_\mathrm{max} = \tsmax$, which corresponds to a local significance of $\sim \tsmaxsigma\,\sigma$. 
For the global significance, we have to account for two sources of trials: the number of GTIs within $t_\mathrm{CC, 95}$ ($N_\mathrm{bins} = 14$ for this SN) and the number of SNe ($N_\mathrm{SN}$) in our sample. 
Thus, the global significance is reduced to $p_\mathrm{global} = 1-(1-(1-p_\mathrm{local})^{N_\mathrm{bins}})^{N_\mathrm{SN}} \sim 0.06~(1.6\,\sigma)$, so we do not find any significant hint for an ALP-triggered \gr burst in our SN sample.  

We proceed by setting limits on $g_{a\gamma}$. 
For the GTI $t_j$ that maximizes the $\mathrm{TS}$ value (denoted with the index $\hat{\jmath}$), we compute the log-likelihood ratio test~\cite{2005NIMPA.551..493R} $\lambda_{i}(g_{a\gamma}) = -2 \sum_k(\ln\mathcal{L}_{i\hat{\jmath}k}(g_{a\gamma}) - \ln\mathcal{L}_{i\hat{\jmath}k}(\hat{g}_{a\gamma}))$ and set one-sided 95\,\% confidence upper limits when $\lambda = 2.71$.
This corresponds to 1 degree of freedom given by the extra parameter of the photon-ALP coupling since the other parameters ($m_a$, explosion time, $B$ field, etc.) are fixed. 
To confirm our statistical procedure and the choice of the GTI, we have conducted a coverage study with simulations including an injected signal. We find that we obtain the right coverage with our chosen procedure of selecting the GTI which corresponds to the highest $\mathrm{TS}$ value (see the SM). 

We show the derived limits SN2017ein in the right panel of Fig.~\ref{fig:lcs} and for all SNe in the SM. 
As expected, the limits do not depend on $m_a$ for low masses and degrade towards higher masses since $P_{a\gamma}$ is reduced. 
The observed limits (red line) compare well against the expected exclusions (black dashed line and green and yellow shaded regions). 
The expected limits
are derived from data by repeating the analysis for all GTIs outside the time interval defined 99\,\% quantile of the posterior $P(t_\mathrm{CC})$.
This effectively provides us with ``off'' regions for the analysis where we do not expect a signal and
has the advantage that the contribution of subthreshold sources is naturally included.

There remains the possibility that an SN was outside the field of view of the LAT during the core collapse. In this case, the derived limits would not be valid. 
We estimate the chance that the LAT indeed observed the SN in one GTI by multiplying the time-averaged expected \gr spectrum with the normalized time-dependent LAT exposure and $P(t_\mathrm{CC})$ and  integrating over the duration of the GTI. 
The total probability that the $i$th SN was observed, $p_{\mathrm{obs},i}$, is then found by summing the individual probabilities for each GTI, which is reported in the SM for each SN.
It is of the order of 10\,\% for each SN. 
The probability that at least one SNe was observed is given by $P(N_\mathrm{SN, obs} > 1) = 1 - \prod_i(1-p_{\mathrm{obs},i})\sim\probtot$ for a spectrum of a 10\,$M_\odot$ progenitor, the NiCo SN explosion model, and the JF12 GMF model. 

\begin{figure}[tbh]
    \centering
    \vspace{10pt}
    \includegraphics[width = .99\linewidth]{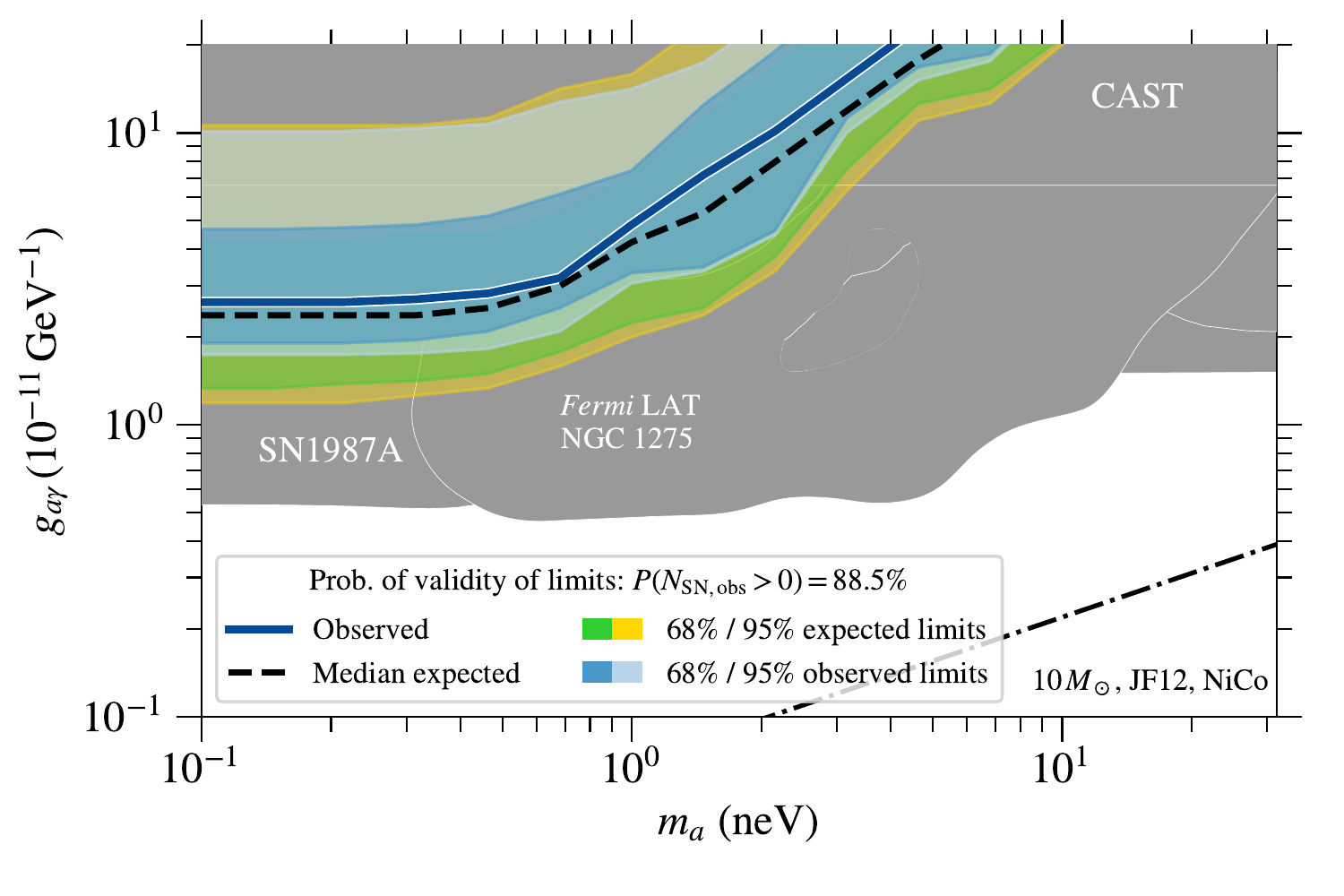}
    \caption{
    Combined observed (blue shaded regions) and expected limits (green and yellow regions) from stacking the likelihoods of different SNe 
    under the assumption that at least one SN was observed during the time of the core collapse.
    The median observed limit is shown as a blue solid line, whereas the median expected limit is shown as a dashed black line. 
    Grey shaded regions show the parameter space excluded by other experiments~\cite{2017NatPh..13..584A, 2015JCAP...02..006P, 2016PhRvL.116p1101A,2013PhRvD..88j2003A,2011PhRvD..84h5001G}. Below the black dash-dotted line, ALPs could constitute the entire dark matter~\cite{arias2012}.
    }
    \label{fig:combined}
\end{figure}

Lastly, we combine the limits from the individual SNe through likelihood stacking, i.e., we sum over $\lambda_i$ with respect to $i$.
In order to decide which SN to include in the sum, we randomly draw a sample of SNe, where each SN is drawn with a success probability of $p_{\mathrm{obs},i}$. 
We repeat this procedure $10^3$ times and show the results in Fig.~\ref{fig:combined}. 
The median combined limit for the cases where at least one SN was drawn from the sample is shown as a blue solid line and the 68\,\% and 95\,\% regions are shown as dark and light blue shaded regions, respectively.
Again, the expected limits (yellow and green shaded regions and black dashed line) agree well with the observed ones.
 
\section{Discussion}
\label{sec:concl}

With the nonobservation of a \gr burst from a sample of 20 extragalactic core-collapse SNe, we are able to rule out photon-ALP couplings 
$g_{a\gamma} > (2.6^{+2.0}_{-0.7})\times 10^{-11}\,\mathrm{GeV}^{-1}$ below $m_a < 3\times10^{-10}\,$eV at 95\,\% confidence under the assumption that all progenitors have a mass of $10\,M_\odot$, that the optical emission can be modeled with the NiCo model in \mosfit, and that the GMF is given by the model of JF12.
The quoted uncertainties represent the 95\,\% spread in the combined limits, 
which derives from the unknown number of SNe in the field of view of the LAT during the core collapse.
These constraints improve previous limits~\cite{2015JCAP...02..006P} by a factor of 2.  
The probability that \FermiLAT observed at least one SN is equal to \probtot.

The dependence of our constraints on the model assumptions are assessed by repeating the analysis for different progenitor masses ($18$ and $40\,M_\odot$), a different GMF model~\cite{2011ApJ...738..192P},
and different models for the evolution of the optical light curves.
The GMF model, the optical light curve model, and a progenitor mass of 18\,$M_\odot$ have only marginal influence on the constraints and the limits end up between $g_{a\gamma} < (2.1$-$3.2) \times 10^{-11}\,\mathrm{GeV}^{-1}$ for $m_a < 3\times 10^{-10}\,$eV.
A comparison of the three tested models for the evolution of the optical light curves reveals that the best fit for the explosion time can differ by several days, highlighting the systematic uncertainties in these fits. 
However, the GTIs in which we search for a signal for the different models indicate no evidence for a signal (the same is true when we assume the literature values for the best-fit explosion times where these are available). 
The largest effect is observed when all progenitors are assumed to have a mass of $40\,M_\odot$. 
In this case, the limits improve to  $g_{a\gamma} < 7.9 \times 10^{-12}\,\mathrm{GeV}^{-1}$ for $m_a \lesssim 1\,$neV. 
Higher progenitor masses are probably more realistic, since the average progenitor mass for a core collapse SN is around $16\,M_\odot$~\cite[e.g.][]{2009JCAP...01..047L} and Wolf-Rayet stars and blue supergiants usually have masses above this average. This also agrees with estimates for the progenitor masses from the optical light curves.
Note that we neglect the correlation between SBO time and progenitor mass~\cite{Kistler:2012as} and opt for the conservative approach to smear the \mosfit predictions with the full range of allowed breakout times.



Our analysis is able to put constraints for masses $m_a \lesssim1\,\mathrm{neV}$ that are within a factor of 5 of the limits obtained from SN1987A for our fiducial set of model parameters (our limits extend to arbitrary low masses since the photon-ALP conversion probability in the Galactic magnetic field is independent of mass for $m_a \lesssim 0.5\,$neV). 
These constraints agree with the expectation that the limits scale as $\sqrt{d}$; for a potential SN in the Andromeda Galaxy ($d=778\,$kpc)  limits of the order of $g_{a\gamma} \sim 2 \times 10^{-12}$ should be possible~\cite{2017PhRvL.118a1103M}.

It should be noted that there has recently been discussions about the validity of the ALP production rate in an SN core~\cite{2017arXiv171206205B} as derived in~\cite{2015JCAP...02..006P}.
While the authors of Ref.~\cite{2015JCAP...02..006P} conservatively only assume an ALP coupling to photons, the exact plasma properties in an SN remain unknown and model dependent and different assumptions could significantly change the ALP production rate~\cite{2017arXiv171206205B}. 
Further theoretical work in this direction would be highly desirable. 

In the future, the SN sample that can be used to search for an ALP-induced \gr burst will grow significantly thanks to currently operating and future optical surveys such as ASAS-SN~\cite{2014ApJ...788...48S}, ZTF~\cite{2019PASP..131a8002B}, 
and the Rubin Observatory~\cite{2009arXiv0912.0201L}. 
We find that with a sample size $\gtrsim 40$ the probability that the LAT has observed one SNe should be $\gtrsim 99\,\%$, which could be achieved within the next five years (see the SM for details).
Running the \textsc{simsurvey} code~\cite{2019JCAP...10..005F}, we estimate that ZTF should observe 26 SNe of type Ib/c per year and 3 (6) of these events within 2 (3) days after the core collapse.
The Rubin Observatory could observe up to $\sim 20$ type Ib/c SNe within a redshift of 0.02 and 1~day after the explosion. 
This provides a strong incentive for continued operations of the \FermiLAT as well as a science case for future gamma-ray satellites such as AMEGO~\cite{2019BAAS...51g.245M}. 

\begin{acknowledgments}

The authors thank Daniel Goldstein for discussions on the number of SNe Ib/c detectable with the Rubin Observatory and Anna Franckowiak as well as Michael Gustafsson for comments and discussions.
TP and MM  acknowledge  support  from the  Fermi  Guest  Investigation grant NNH17ZDA001N.
TP acknowledges the financial support from the Slovenian Research Agency (grants I0-0033, P1-0031, J1-8136 and Z1-1853). This work was supported by a collaborative visit funded by the Slovenian Research Agency (ARRS, travel grant number BI-US/18-20-017) and by the Cosmology and Astroparticle Student and Postdoc Exchange Network (CASPEN).

The \textit{Fermi}-LAT Collaboration acknowledges support for LAT development, operation and data analysis from NASA and DOE (United States), CEA/Irfu and IN2P3/CNRS (France), ASI and INFN (Italy), MEXT, KEK, and JAXA (Japan), and the K.A.~Wallenberg Foundation, the Swedish Research Council and the National Space Board (Sweden). Science analysis support in the operations phase from INAF (Italy) and CNES (France) is also gratefully acknowledged. This work performed in part under DOE Contract DE-AC02-76SF00515.

\end{acknowledgments}

\bibliographystyle{JHEP}
\bibliography{references}

\ifArxiv

\ifx\ifArxiv\undefined

\newif\ifArxiv
\Arxivtrue

\ifArxiv
\definecolor{lime}{HTML}{A6CE39}
\DeclareRobustCommand{\orcidicon}{
	\begin{tikzpicture}
	\draw[lime, fill=lime] (0,0) 
	circle [radius=0.16] 
	node[white] {{\fontfamily{qag}\selectfont \tiny ID}};
	\draw[white, fill=white] (-0.0625,0.095) 
	circle [radius=0.007];
	\end{tikzpicture}
	\hspace{-2mm}
}
\foreach \x in {A, ..., Z}{\expandafter\xdef\csname orcid\x\endcsname{\noexpand\href{https://orcid.org/\csname orcidauthor\x\endcsname}
			{\noexpand\orcidicon}}
}
\newcommand{\orcidauthorA}{0000-0002-0738-7581} 
\newcommand{\orcidauthorB}{0000-0003-4743-1679} 
\newcommand{\figsens}{15}

\else

\newcommand{\figsens}{\ref{fig:sensitivity}}

\fi

\fi

\ifArxiv


\else

\documentclass[%
 reprint,
nofootinbib,
 amsmath,amssymb,
 aps,
 prd,
]{revtex4-2}

\usepackage{graphicx}
\usepackage{dcolumn}
\usepackage{bm}
\usepackage[colorlinks = true,
            linkcolor = black,
            urlcolor  = blue,
            citecolor = blue,
            anchorcolor = black]{hyperref}
\usepackage{aas_macros}
\usepackage{xspace}
\usepackage{multirow}


\newcommand{\mosfit}{\textsc{MOSFiT}\xspace}
\newcommand{\gr}{$\gamma$-ray\xspace}
\newcommand{\Grs}{$\gamma$~rays\xspace}
\newcommand{\fermi}{\textit{Fermi}\xspace}
\newcommand{\fermiLAT}{\textit{Fermi}-LAT\xspace}
\newcommand{\FermiLAT}{\textit{Fermi}~LAT\xspace}
\newcommand{\probtot}{88.5\,\%}
\newcommand{\tsmax}{12.4}
\newcommand{\tsmaxsigma}{3.5}

\begin{document}

\title{Supplemental Material --- Search for axionlike-particle induced prompt $\gamma$-ray emission from extragalactic core-collapse supernovae with the \emph{Fermi} Large Area Telescope}

\author{Manuel Meyer}
 \email{manuel.e.meyer@fau.de}
\affiliation{%
Erlangen Centre for Astroparticle Physics, University of Erlangen-Nuremberg, Erwin-Rommel-Str. 1, 91058 Erlangen, Germany
}
\affiliation{
W. W. Hansen Experimental Physics Laboratory, Kavli Institute for
Particle Astrophysics and Cosmology, Department of Physics and SLAC National Accelerator Laboratory, Stanford University, Stanford, CA
94305, USA
}%

\collaboration{\emph{Fermi}-LAT Collaboration}

\author{Tanja Petrushevska}
\email{tanja.petrushevska@ung.si}
\affiliation{
 Centre for Astrophysics and Cosmology, University of Nova Gorica, Vipavska 11c, 5270 Ajdov\v{s}\u{c}ina, Slovenia
}%

\date{\today}

\fi

\ifArxiv

\onecolumngrid
\section*{Supplemental Material}
\appendix

\else

\maketitle
\onecolumngrid

\fi

\section{SN sample}

Our selected sample of SNe is listed in Tab.~\ref{tab:gr-result} together with the results for the time interval of the core collapse and from the \gr analysis. 

\begin{table*}[h]
\centering
\begin{tabular}{l|cccccccc}
\hline
\hline
SN & R.A. (deg) & Dec. (deg) & Redshift & $t_\mathrm{CC}$ (MJD) & $N_\mathrm{bins}$ & $\mathrm{TS}_\mathrm{max}$ & $p_\mathrm{obs}$ & Ref. \\
\hline

SN2009bb & $157.891$ & $-39.958$ & 0.0104 & $54909.20^{+0.55}_{-0.53}$& 14 & 2.59 & 0.094 &
\cite{2011ApJ...728...14P} \\
SN2009iz & $40.564$ & $42.397$ & 0.014 & $55087.06^{+1.62}_{-2.79}$& 42 & 0.00 & 0.099 &
\cite{2014ApJS..213...19B} \\
SN2009jf & $346.221$ & $12.333$ & 0.0079 & $55099.03^{+0.45}_{-0.55}$& 9 & 7.03 & 0.063 &  
\cite{2011MNRAS.413.2583S,2011MNRAS.416.3138V} \\
SN2010bh & $107.632$ & $-56.256$ & 0.0593 & $55271.44^{+0.15}_{-0.13}$& 3 & 2.68 & 0.107 &
\cite{2011ApJ...740...41C} \\
SN2010et & $259.225$ & $31.564$ & 0.023 & $55344.90^{+1.18}_{-1.22}$& 35 & 8.07 & 0.099 & 
\cite{2012ApJ...755..161K} \\
SN2011bm & $194.225$ & $22.375$ & 0.022 & $55639.70^{+0.78}_{-0.92}$& 14 & 12.43 & 0.072 & 
\cite{2012ApJ...749L..28V} \\
SN2012P & $224.996$ & $1.890$ & 0.004506 & $55929.84^{+1.64}_{-2.09}$& 55 & 5.31 & 0.108 &
\cite{2016AA...593A..68F} \\
SN2012ap & $75.057$ & $-3.348$ & 0.01224 & $55960.066^{+1.510}_{-1.727}$& 25 & 1.30 & 0.152 &
\cite{2015ApJ...799...51M} \\
PTF12gzk & $333.173$ & $0.512$ & 0.01377 & $56131.49^{+0.13}_{-0.13}$& 3 & 0.35 & 0.127 &
\cite{2012ApJ...760L..33B} \\
iPTF13bvn & $225.001$ & $1.881$ & 0.00449 & $56457.20^{+0.20}_{-0.21}$& 5 & 0.00 & 0.090 & 
\cite{2016AA...593A..68F} \\
SN2013ge & $158.702$ & $21.662$ & 0.004356 & $56595.54^{+1.29}_{-0.66}$& 31 & 5.55 & 0.138 &
\cite{2016ApJ...821...57D} \\
SN2014L & $184.703$ & $14.412$ & 0.008029 & $56681.33^{+0.31}_{-0.30}$& 9 & 3.01 & 0.079 &
\cite{2018ApJ...863..109Z} \\
SN2014ad & $179.435$ & $-10.171$ & 0.0057 & $56724.14^{+1.54}_{-1.53}$& 46 & 2.78 & 0.054 &
\cite{2018MNRAS.475.2591S} \\
SN2015ap & $31.306$ & $6.102$ & 0.01138 & $57271.22^{+0.58}_{-0.93}$& 17 & 2.67 & 0.089 &
\cite{2014ApSS.354...89B} \\
PTF15dtg & $37.584$ & $37.235$ & 0.0524 & $57322.52^{+2.29}_{-3.96}$& 56 & 1.96 & 0.113 & 
\cite{2016AA...592A..89T} \\
SN2016bau & $170.246$ & $53.174$ & 0.003856 & $57459.08^{+0.64}_{-0.84}$& 10 & 0.00 & 0.041 & 
(a) \\
SN2016blz & $235.122$ & $0.910$ & 0.01173 & $57480.94^{+2.38}_{-3.11}$& 83 & 7.38 & 0.142 &
(a) \\
SN2016coi & $329.767$ & $18.186$ & 0.003646 & $57530.58^{+1.51}_{-1.43}$& 40 & 2.55 & 0.113 & 
\cite{2018MNRAS.478.4162P} \\
SN2017ein & $178.222$ & $44.124$ & 0.002699 & $57896.36^{+0.49}_{-0.58}$& 10 & 0.69 & 0.146 &
\cite{2018ApJ...860...90V} \\
SN2017fwm & $288.217$ & $-60.383$ & 0.015557 & $57965.11^{+0.15}_{-0.25}$& 4 & 0.14 & 0.041 &
(a,b) \\

\hline


\hline
\end{tabular}
\caption{\label{tab:gr-result}
The sample of SNe Ib/c studied in this work, together with the estimate of the time of the core collapse $t_\mathrm{CC}$, the number of GTIs within $t_\mathrm{CC, 95}$, the highest $\mathrm{TS}$ value found for these GTIs, as well as the probability that \FermiLAT observed the core collapse, $p_\mathrm{obs}$. 
Further notes:
\newline (a) Photometric data not in any publication yet, downloaded from the OSC.
\newline (b) \url{http://gsaweb.ast.cam.ac.uk/alerts}}
\end{table*}

\section{Fits to optical SN light curves}

The prevailing procedure to derive the estimates of the explosion parameters, such as the ejecta mass ($M_{\rm ej}$), the $^{56}$Ni mass ($M_{\rm Ni}$) and kinetic energy ($E_k$), is to fit the main peak of the bolometric light curve and the photospheric velocity using the simple analytical Arnett model \cite{1982ApJ...253..785A}, in which the light curve is powered by the $^{56}$Ni radioactivity and/or further slight modifications of it \cite[e.g.,][]{2008MNRAS.383.1485V, 2015A&A...574A..60T}. 
The model assumes homologous expansion of the ejecta, spherical symmetry, and that all radioactive $^{56}$Ni is located in the centre. It adopts a constant optical opacity $\kappa$ (which is usually assumed in the literature to be $0.06$-$0.07$ $\rm cm^{2}g^{-1}$), a small initial radius before explosion, and optically thick ejecta. The characteristic photospheric velocity is an input to the model, and some appropriate value is assumed, or measured from spectra. The peak luminosity is directly correlated with $M_{\rm Ni}$ and the width of the light curve peak is correlated to $M_{\rm ej}$. 

The time of the explosion is usually determined either from a fit using a simple equation of the expanding fireball, where the early light curve follows a polynomial, or from considerations of nondetections of the SN in the data before the first detection date. 
There are several limitations in making the aforementioned assumptions, e.g., it is observed that constant opacity and spherical symmetry are not quite representative for SNe Ib/c.
For this reason, the best-fit parameters suffer large systematic uncertainties which are not always reported in the literature. 

In our case, we determine the marginalized posterior for the explosion time (i.e., the onset of the optical emission) using \mosfit. 
The implemented models are described in detail in Refs.~\cite{2017ApJ...849...70V,2017ApJ...850...55N} and we provide a short summary below. 

The bolometric luminosity of the SN is given by 
\begin{equation}
    L_\mathrm{obs}(t) = e^{-(t/t_\mathrm{diff})^2}\left(1-e^{-At^{-2}}\right)
    \int\limits_0^t2L_\mathrm{in}(t')\frac{t'}{t_\mathrm{diff}}
    e^{-(t/t_\mathrm{diff})^2}
    \frac{\mathrm{d}t'}{t_\mathrm{diff}},
\end{equation}{}
with the diffusion time
\begin{equation}
    t_\mathrm{diff} = 
    \left(\frac{2\kappa M_\mathrm{ej}}{\beta c \nu_\mathrm{ej}}\right)^{1/2}
\end{equation}{}
and leakage parameter
\begin{equation}
    A = \frac{3\kappa_\gamma M_\mathrm{ej}}{4\pi \nu_\mathrm{ej}^2}
\end{equation}{}
where $\kappa$ ($\kappa_\gamma$) is the (high-energy) photon opacity, $M_\mathrm{ej}$ is the ejecta mass, $\nu_\mathrm{ej}$ is the ejecta velocity, and $\beta = 4\pi^3 / 9$ is a geometric correction factor~\cite{1982ApJ...253..785A}. 
For the engine $L_\mathrm{in}(t)$, we either assume that the SNe are powered by Nickel and Cobalt (NiCo)  decay~\cite{1994ApJS...92..527N}, 
\begin{equation}
    L_\mathrm{in}(t) = M_\mathrm{ej} f_\mathrm{Ni} \left(L_\mathrm{Ni}e^{-(t - t_\mathrm{exp})/\tau_\mathrm{Ni}} + L_\mathrm{Co}e^{-(t - t_\mathrm{exp})/\tau_\mathrm{Co}}  \right),
\end{equation}

where $L_\mathrm{Ni} = 6.45\times10^{43}\mathrm{erg}\,\mathrm{s}^{-1}$ and $L_\mathrm{Co} = 1.45\times10^{43}\mathrm{erg}\,\mathrm{s}^{-1}$, with the lifetimes $\tau_\mathrm{Ni} = 8.8\,$days and $\tau_\mathrm{Co} = 111.3$\,days, 
and free parameters $M_\mathrm{ej}$, $f_\mathrm{Ni}$ and $t_\mathrm{exp}$. The parameter $f_\mathrm{Ni}$ is the Nickel mass fraction and $t_\mathrm{exp}$ denotes the time of the onset of the optical emission. 
For the NiCo model with an addtional shock breakout component (SBO+NiCo model), we follow Refs.~\cite{2007ApJ...667..351W,cowen2010} 
and add an additional SBO component, $L_\mathrm{SBO}$ to the observed luminosity $L_\mathrm{obs}$,

\begin{equation}
    L_\mathrm{SBO}(t) = \frac{\L_\mathrm{scale}(t - t_\mathrm{exp})^{1.6}}{\exp\left(\alpha\sqrt{t-t_\mathrm{exp}}\right) - 1},
\end{equation}

which should describe the SBO phase and introduces two additional free parameters, $\alpha$ and $\L_\mathrm{scale}$. The full model in this case is then given by $L_\mathrm{SBO} + L_\mathrm{obs}$.

Alternatively, we use a generic exponential rise followed by a power-law decay (exppow), 
\begin{equation}
    L_\mathrm{in}(t) = \L_\mathrm{scale}\left(1 - e^{-(t - t_\mathrm{exp})/t_\mathrm{peak}}\right)^\alpha \left(\frac{t - t_\mathrm{exp}}{t_\mathrm{peak}}\right)^{-\beta},
\end{equation}

with free parameters $\L_\mathrm{scale}, t_\mathrm{exp}, t_\mathrm{peak}, \alpha, \beta$.
The opacities $\kappa$ and $\kappa_\gamma$ are left as additional free parameters in both models.
In order to convert the bolometric luminosity into the luminosity in a certain optical wavelength band, we need to assume a spectral energy distribution (SED). 
Here, we follow the default \mosfit settings and use a simple black body SED, which, in turn, requires a model for the photosphere. 
For the NiCo and SBO+NiCo engines, we use the \texttt{photosphere.temperature\_floor} module in \mosfit, which introduces an additional parameter for the temperature floor, $T_\mathrm{min}$~\cite[see][for further details]{2017ApJ...850...55N}. 
For the exppow model, we use instead the \texttt{photosphere.densecore} module, which does not introduce additional parameters.
The photosphere radius and temperature, which enter into the black body SED, depend on $L_\mathrm{obs}(t)$ and $\nu_\mathrm{ej}$.
For the latter, it is assumed that it is equal to the velocity of the expanding photosphere. 

For all free model parameters, flat priors in linear or logarithmic space  are assumed. 
If published light curves are already corrected for host and/or Galactic extinction, the corresponding $E(B-V)$ values and host column densities $n_\mathrm{Host}$ are fixed to zero, otherwise they are treated as additional free parameters. 
Since we are mainly interested in the early time behavior of the SN light curve and a good estimation of the explosion time, we restrict ourselves to data within the first 50 days after the discovery. 
We have tested that using shorter or longer intervals does not significantly change the best-fit value of the explosion time. 

The fit is performed using an MCMC implementation of the nested sampler \textsc{dynesty}\footnote{\url{https://dynesty.readthedocs.io}}~\cite{Higson2019}.
The posterior is sampled and the volume integral is performed which gives the evidence $Z$. 
As a convergence criterion, we choose the default \mosfit setting of $\Delta\log Z = 0.02$ (see the \mosfit and \textsc{dynesty} documentations for further details).

We use the data available from the OSC. However, we noticed that for some SNe data points are missing in the database, especially upper limits on the flux at times before the SN was detected. 
This is likely due to the fact that these upper limits are not reported in some of the tables provided in the respective publications. 
We have augmented our data with these limits, which we summarize in Tab.~\ref{tab:missingdata}.

\begin{table}
    \centering
    \begin{tabular}{l|cccc}
    \hline
    \hline
    Source name & Time (MJD) & Band & UL magnitude & Reference \\
    \hline
SN2009jf        &        55098.11 & B               & 19.5            & \cite{2011MNRAS.416.3138V} \\
SN2009jf        &        55098.11 & V               & 19.0            & \cite{2011MNRAS.416.3138V} \\
SN2009jf        &        55098.11 & R               & 18.5            & \cite{2011MNRAS.416.3138V} \\
SN2010bh        &        55271.53 & XRF$^\mathrm{a}$             & ---             & \cite{2011ApJ...740...41C} \\
SN2011bm        &        55643.00 & P48-R           & 20.8            & \cite{2011ATel.3288....1G} \\
SN2012P         &        55932.00 & P48-R           & 18.9            & \cite{2012ATel.3881....1A} \\
SN2012ap        &        55962.21 & KAIT-R          & 18.7            & \cite{2015ApJ...799...51M} \\
PTF12gzk        &        56128.00 & P48-R           & 21.5            & \cite{2012ApJ...760L..33B} \\
iPTF13bvn       &        56458.21 & P48-R           & 21              & \cite{2013ATel.5137....1C} \\
SN2013ge        &        56597.00 & R               & 19              & \cite{2016ApJ...821...57D} \\
iPTF15dtg       &        57332.43 & P48-g           & 20.46           & \cite{2016AA...592A..89T} \\
iPTF15dtg       &        57332.46 & P48-g           & 20.16           & \cite{2016AA...592A..89T} \\
SN2017ein       &        57896.77 & R               & 19.9            & \cite{2017ATel10481....1I} \\
SN2017ein       &        57897.63 & R               & 18.4            & \cite{2017ATel10481....1I} \\
\hline
    \end{tabular}
    \caption{Additional upper limit (UL) values on flux magnitudes not included in the OSC but used here. Notes: \newline
    a. SN2010bh is associated with the X-ray flash XRF~100316D. We limit the prior on $t_\mathrm{exp}$, so that $t_\mathrm{exp}$ needs to be between the time of the XRF and the first detection in optical bands.}
    \label{tab:missingdata}
\end{table}{}

\begin{figure*}
\centering
\includegraphics[width = 0.95\linewidth]{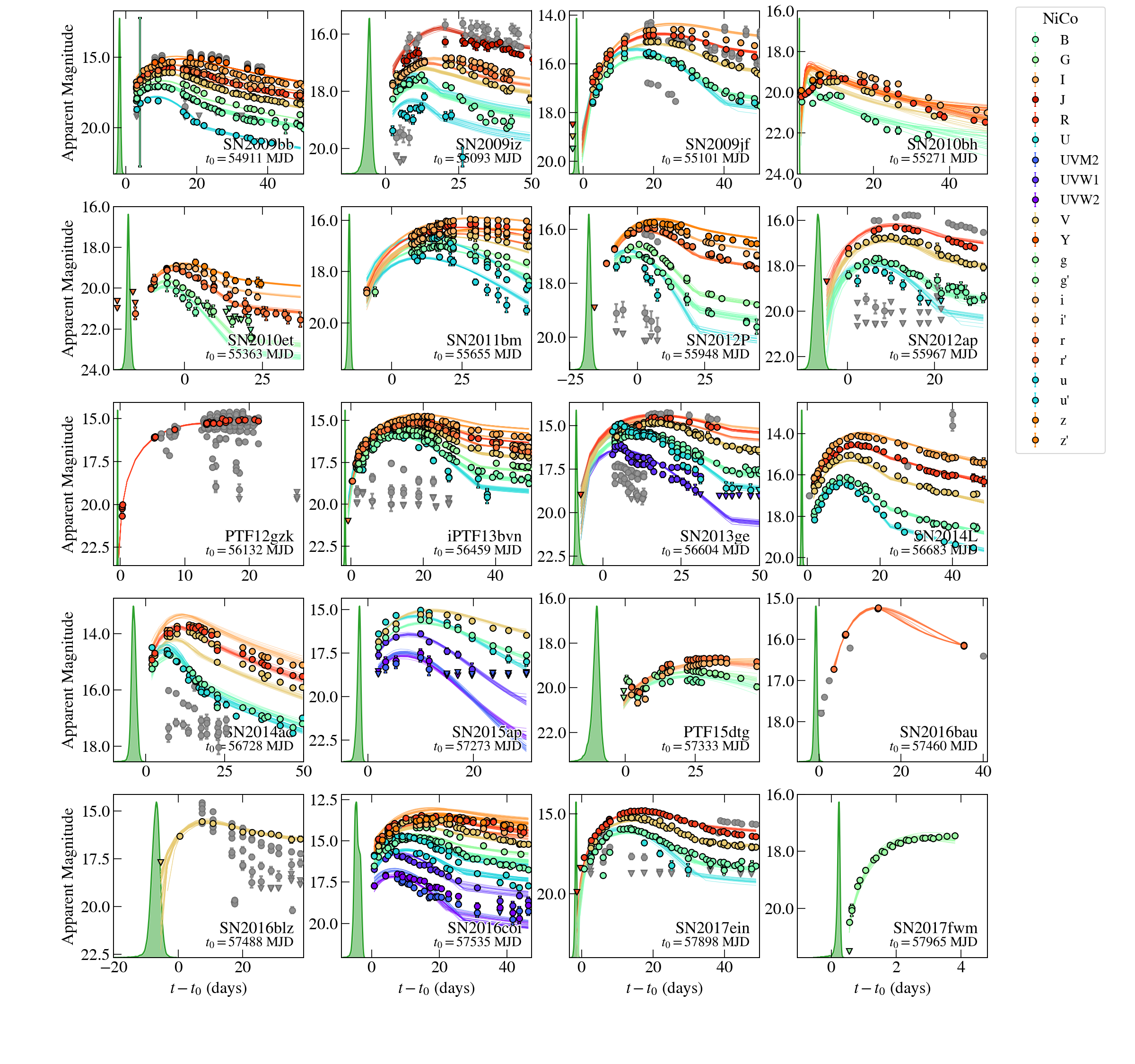}
\caption{Optical SN light curves with NiCo model realizations randomly drawn from the posterior. Data points shown in grey are not included in the fit. The green shaded curve displays the marginalized posterior of the explosion time.  \label{fig:mosfit-nico-lcs}}
\end{figure*}

\begin{figure*}
\centering
\includegraphics[width = 0.95\linewidth]{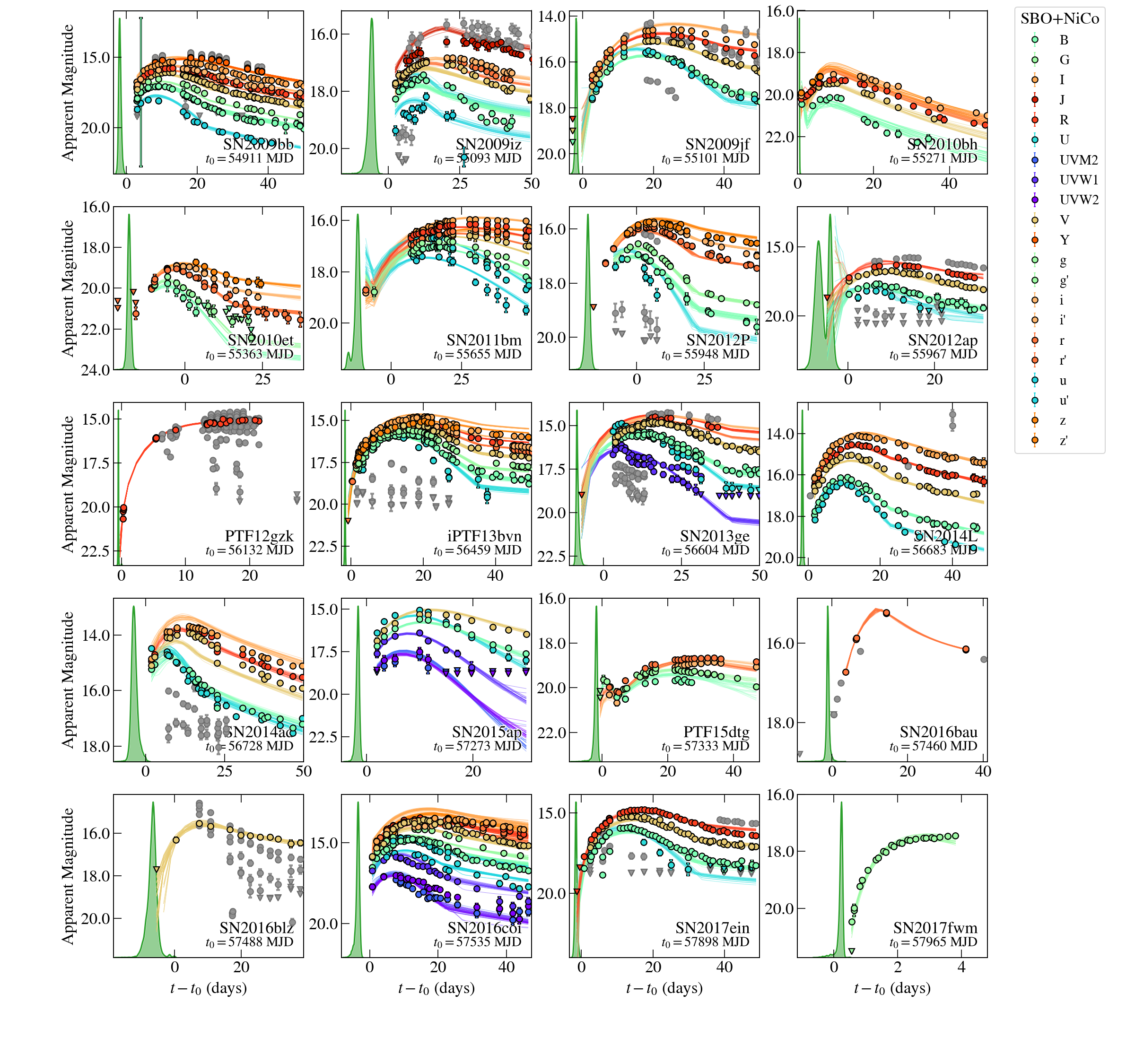}
\caption{
Same as Fig.~\ref{fig:mosfit-nico-lcs} but for the SBO+NiCo model. The SBO component is clearly visible for SN2010bh, SN2011bm, and PTF15dtg.
\label{fig:mosfit-sbonico-lcs}}
\end{figure*}

\begin{figure*}
\centering
\includegraphics[width = 0.95\linewidth]{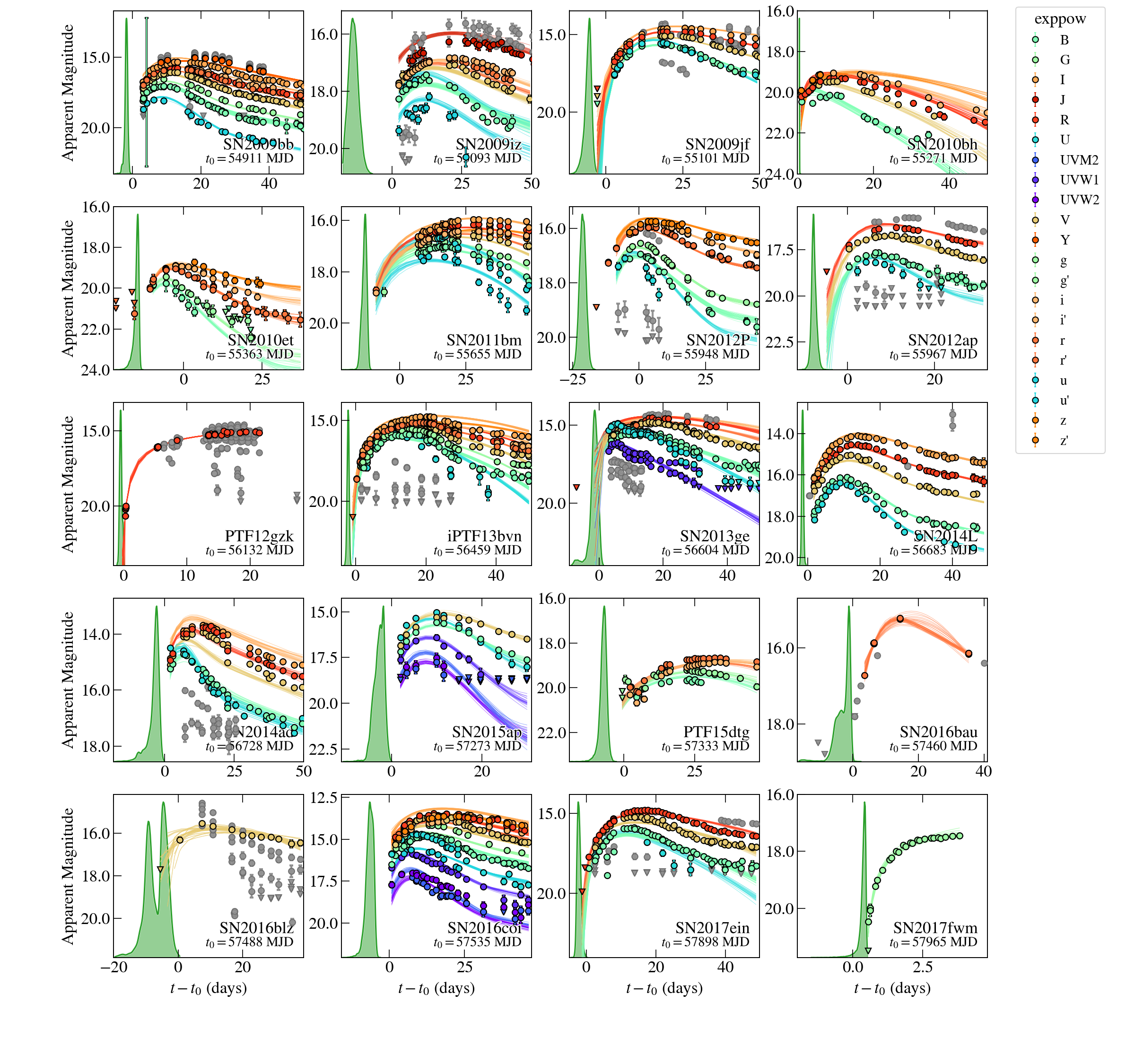}
\caption{Same as Fig.~\ref{fig:mosfit-nico-lcs} but for the exppow model.\label{fig:mosfit-exppow-lcs}}
\end{figure*}

\begin{figure*}
\centering
\includegraphics[width = 0.95\linewidth]{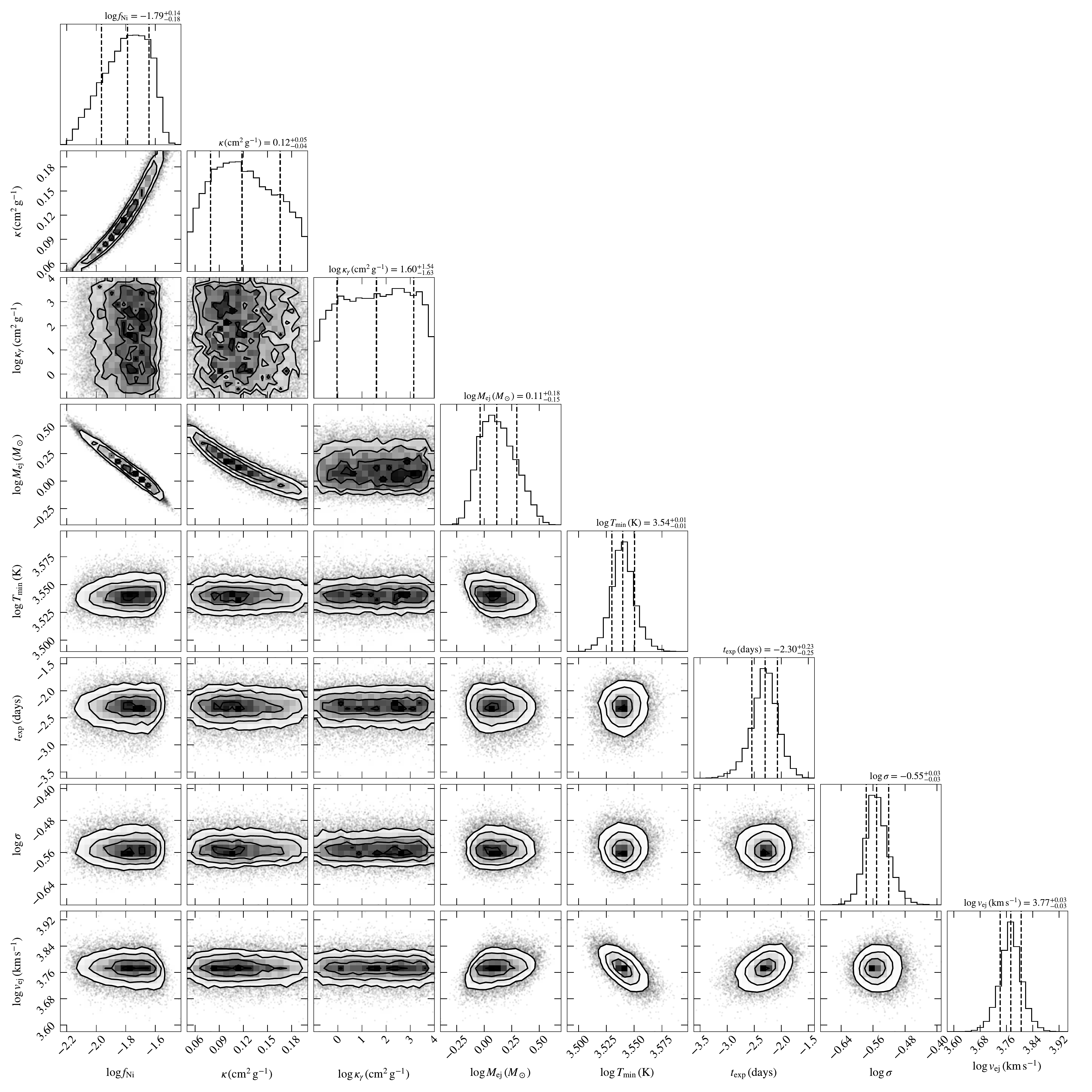}
\caption{Corner plot for SN2017ein and the NiCo model. Except for the opacities $\kappa$ and $\kappa_\gamma$, all model parameters are well constrained in the MCMC method and the posterior distributions are well sampled. \label{fig:mosfit-default-corner}}
\end{figure*}

The best-fit parameters for the NiCo, SBO+NiCo, and exppow models are presented in Tabs.~\ref{tab:mosfit-nico}, \ref{tab:mosfit-sbonico}, and \ref{tab:mosfit-exppow}, respectively, and the light curves together with realizations of the posterior are shown in Figs.~\ref{fig:mosfit-nico-lcs}, \ref{fig:mosfit-sbonico-lcs}, and \ref{fig:mosfit-exppow-lcs}. 
As an example, we also show a \textit{corner} plot of the nested MCMC sampling for the NiCo model and SN2017ein in Fig.~\ref{fig:mosfit-default-corner}. 
In this plot, one notices an additional parameter, $\sigma$, which is an additional variance parameter. 
It is added to each uncertainty of the measured magnitude so that the reduced $\chi^2$ approaches 1~\cite{2017ApJ...835...64G}, 
see also Eq.~1 in Ref.~\cite{2017ApJ...850...55N}.
It can thus be regarded as a quantification of the model uncertainty.\footnote{See also \url{https://mosfit.readthedocs.io/en/latest/error.html} for further details.}
The best-fit values for $\sigma$ are also provided in Tabs.~\ref{tab:mosfit-nico}, \ref{tab:mosfit-sbonico}, and~\ref{tab:mosfit-exppow}.

Inspecting the light curves it becomes clear that the overall time evolution of the SNe light curves is well captured in most cases. 
However, the models often overshoot the measurements in the I band, which indicates that the absorption of emission is not accurately modeled. 
As this happens predominantly at later times, the effect on the best-fit explosion times should not be too strong. 
Furthermore, it becomes obvious that SN light curves with early detections are much better suited to provide tight constraints on the explosion times. 

\begin{table*}
\centering
\begin{scriptsize}
\begin{tabular}{l|ccccccccc}
\hline
\hline
Source name & $\log\, M_{\rm ej}\,(M_\odot)$ & $\log\, f_{\rm Ni}$ & $\kappa\,({\rm cm}^{2}\,{\rm g}^{-1})$ & $\log\, \kappa_\gamma\,({\rm cm}^{2}\,{\rm g}^{-1})$ & $\log\, v_{\rm ej}\,({\rm km\,s}^{-1})$ & $\log\, T_{\min}\,{\rm (K)}$ & $\log\, \sigma$ & $t_{\rm exp}\,{\rm (days)}$ & $t_0$ (MJD) \\
\hline
SN2009bb & $0.17_{-0.06}^{+0.07}$ & $-0.87_{+0.12}^{-0.15}$ & $0.19_{-0.01}^{+0.01}$ & $-0.98_{+0.02}^{-0.01}$ & $4.38_{-0.03}^{+0.03}$ & $3.60_{-0.03}^{+0.02}$ & $-0.71_{+0.02}^{-0.02}$ & $-4.68_{+0.26}^{-0.25}$ &  54914.00  \\\\
SN2009iz & $0.26_{-0.15}^{+0.18}$ & $-1.19_{+0.14}^{-0.18}$ & $0.12_{-0.04}^{+0.05}$ & $1.34_{-1.44}^{+1.64}$ & $3.94_{-0.05}^{+0.04}$ & $3.74_{-0.01}^{+0.01}$ & $-0.63_{+0.03}^{-0.03}$ & $-8.26_{+0.79}^{-1.14}$ &  55095.43  \\\\
SN2009jf & $0.39_{-0.12}^{+0.19}$ & $-1.14_{+0.11}^{-0.20}$ & $0.13_{-0.04}^{+0.04}$ & $1.60_{-1.64}^{+1.52}$ & $3.95_{-0.02}^{+0.02}$ & $3.61_{-0.01}^{+0.01}$ & $-0.77_{+0.04}^{-0.03}$ & $-4.53_{+0.20}^{-0.23}$ &  55103.67  \\\\
SN2010bh & $-0.93_{+0.10}^{-0.05}$ & $-0.13_{+0.07}^{-0.11}$ & $0.06_{-0.01}^{+0.02}$ & $1.57_{-0.31}^{+1.37}$ & $4.97_{-0.05}^{+0.02}$ & $3.74_{-0.03}^{+0.03}$ & $-0.24_{+0.05}^{-0.04}$ & $-0.45_{+0.02}^{-0.01}$ &  55272.00  \\\\
SN2010et & $-0.19_{+0.21}^{-0.15}$ & $-1.34_{+0.15}^{-0.19}$ & $0.12_{-0.05}^{+0.05}$ & $1.54_{-1.60}^{+1.62}$ & $3.89_{-0.02}^{+0.02}$ & $3.51_{-0.01}^{+0.01}$ & $-2.06_{+0.50}^{-0.54}$ & $-8.17_{+0.54}^{-0.51}$ &  55353.19  \\\\
SN2011bm & $0.67_{-0.15}^{+0.18}$ & $-0.86_{+0.15}^{-0.18}$ & $0.12_{-0.04}^{+0.05}$ & $1.42_{-1.50}^{+1.63}$ & $3.87_{-0.01}^{+0.01}$ & $3.36_{-0.22}^{+0.21}$ & $-0.91_{+0.03}^{-0.03}$ & $-6.23_{+0.38}^{-0.39}$ &  55646.03  \\\\
SN2012P & $0.21_{-0.14}^{+0.18}$ & $-1.75_{+0.14}^{-0.18}$ & $0.12_{-0.04}^{+0.05}$ & $1.59_{-1.62}^{+1.53}$ & $3.73_{-0.02}^{+0.02}$ & $3.54_{-0.00}^{+0.00}$ & $-0.87_{+0.04}^{-0.03}$ & $-9.53_{+0.76}^{-0.84}$ &  55939.48  \\\\
SN2012ap & $0.66_{-0.14}^{+0.18}$ & $-1.43_{+0.13}^{-0.18}$ & $0.13_{-0.04}^{+0.04}$ & $1.79_{-1.67}^{+1.42}$ & $4.33_{-0.03}^{+0.03}$ & $3.54_{-0.01}^{+0.01}$ & $-0.84_{+0.06}^{-0.05}$ & $-7.08_{+0.76}^{-0.86}$ &  55967.25  \\\\
PTF12gzk & $1.26_{-0.14}^{+0.17}$ & $-1.38_{+0.13}^{-0.16}$ & $0.12_{-0.04}^{+0.04}$ & $1.23_{-1.26}^{+1.53}$ & $4.98_{-0.03}^{+0.02}$ & $3.54_{-0.01}^{+0.01}$ & $-1.38_{+0.11}^{-0.08}$ & $-0.66_{+0.02}^{-0.02}$ &  56132.26  \\\\
iPTF13bvn & $-0.04_{-0.12}^{+0.21}$ & $-1.36_{+0.14}^{-0.18}$ & $0.12_{-0.04}^{+0.05}$ & $1.55_{-1.62}^{+1.54}$ & $3.82_{-0.01}^{+0.01}$ & $3.56_{-0.01}^{+0.00}$ & $-0.66_{+0.02}^{-0.02}$ & $-1.92_{+0.07}^{-0.08}$ &  56459.24  \\\\
SN2013ge & $0.53_{-0.14}^{+0.18}$ & $-1.49_{+0.14}^{-0.18}$ & $0.12_{-0.04}^{+0.05}$ & $1.53_{-1.59}^{+1.59}$ & $3.91_{-0.02}^{+0.02}$ & $3.54_{-0.01}^{+0.01}$ & $-0.62_{+0.03}^{-0.02}$ & $-11.36_{+0.52}^{-0.30}$ &  56607.00  \\\\
SN2014L & $0.60_{-0.05}^{+0.06}$ & $-1.18_{+0.04}^{-0.06}$ & $0.17_{-0.03}^{+0.02}$ & $-0.92_{+0.07}^{-0.05}$ & $4.54_{-0.02}^{+0.01}$ & $3.53_{-0.00}^{+0.00}$ & $-0.81_{+0.03}^{-0.03}$ & $-3.28_{+0.13}^{-0.13}$ &  56684.72  \\\\
SN2014ad & $0.07_{-0.08}^{+0.08}$ & $-0.81_{+0.07}^{-0.08}$ & $0.17_{-0.03}^{+0.02}$ & $-0.89_{+0.12}^{-0.07}$ & $4.29_{-0.04}^{+0.06}$ & $3.66_{-0.01}^{+0.01}$ & $-0.55_{+0.04}^{-0.03}$ & $-5.65_{+0.76}^{-0.69}$ &  56729.91  \\\\
SN2015ap & $0.25_{-0.12}^{+0.16}$ & $-0.24_{+0.15}^{-0.22}$ & $0.10_{-0.03}^{+0.05}$ & $1.73_{-1.58}^{+1.48}$ & $4.45_{-0.03}^{+0.03}$ & $3.41_{-0.25}^{+0.24}$ & $-0.80_{+0.07}^{-0.05}$ & $-3.64_{+0.29}^{-0.40}$ &  57274.97  \\\\
PTF15dtg & $0.62_{-0.17}^{+0.21}$ & $-0.92_{+0.15}^{-0.20}$ & $0.12_{-0.04}^{+0.05}$ & $1.43_{-1.54}^{+1.63}$ & $3.73_{-0.04}^{+0.04}$ & $3.44_{-0.26}^{+0.26}$ & $-0.66_{+0.05}^{-0.04}$ & $-10.82_{+1.12}^{-1.56}$ &  57333.43  \\\\
SN2016bau & $-0.51_{+0.23}^{-0.18}$ & $-1.01_{+0.21}^{-0.25}$ & $0.10_{-0.03}^{+0.04}$ & $2.45_{-3.41}^{-2.48}$ & $3.58_{-0.02}^{+0.14}$ & $3.55_{-0.38}^{+0.31}$ & $-1.72_{+0.23}^{-0.18}$ & $-4.41_{+0.29}^{-0.36}$ &  57463.61  \\\\
SN2016blz & $0.32_{-0.55}^{+0.41}$ & $-0.65_{+0.43}^{-0.44}$ & $0.12_{-0.04}^{+0.04}$ & $1.29_{-1.50}^{+1.65}$ & $4.19_{-0.26}^{+0.40}$ & $3.87_{-0.14}^{+0.18}$ & $-1.59_{+0.50}^{-0.54}$ & $-7.41_{+1.18}^{-1.40}$ &  57488.44  \\\\
SN2016coi & $0.28_{-0.30}^{+0.15}$ & $-0.77_{+0.29}^{-0.39}$ & $0.18_{-0.03}^{+0.01}$ & $-0.96_{+0.10}^{-0.02}$ & $4.06_{-0.03}^{+0.04}$ & $3.69_{-0.05}^{+0.07}$ & $-0.45_{+0.02}^{-0.02}$ & $-5.01_{+0.90}^{-0.77}$ &  57535.77  \\\\
SN2017ein & $0.12_{-0.15}^{+0.18}$ & $-1.79_{+0.14}^{-0.18}$ & $0.12_{-0.04}^{+0.05}$ & $1.52_{-1.60}^{+1.60}$ & $3.77_{-0.03}^{+0.03}$ & $3.54_{-0.01}^{+0.01}$ & $-0.55_{+0.03}^{-0.03}$ & $-2.30_{+0.23}^{-0.25}$ &  57898.77  \\\\
SN2017fwm & $-0.84_{+0.15}^{-0.10}$ & $-0.62_{+0.19}^{-0.16}$ & $0.08_{-0.02}^{+0.03}$ & $1.62_{-1.39}^{+1.46}$ & $4.86_{-0.11}^{+0.09}$ & $3.84_{-0.41}^{+0.23}$ & $-0.71_{+0.11}^{-0.07}$ & $-0.33_{+0.04}^{-0.07}$ &  57965.56  \\\\

\hline
\end{tabular}
\end{scriptsize}
\caption{Best-fit parameters and 68\,\% uncertainties for the NiCo model. \label{tab:mosfit-nico}}
\end{table*}

\begin{turnpage}
\begin{table*}
\centering
\begin{scriptsize}
\begin{tabular}{l|ccccccccccc}
\hline
\hline
Source name & $\log\, M_{\rm ej}\,(M_\odot)$ & $f_{\rm Ni}$ & $\kappa\,({\rm cm}^{2}\,{\rm g}^{-1})$ & $\log\, \kappa_\gamma\,({\rm cm}^{2}\,{\rm g}^{-1})$ & $\log\, v_{\rm ej}\,({\rm km\,s}^{-1})$ & $\log\, T_{\min}\,{\rm (K)}$ & $\alpha$ & $\log\, \L_{\rm scale}$ & $\log\, \sigma$ & $t_{\rm exp}\,{\rm (days)}$ & $t_0$ (MJD) \\
\hline
SN2009bb & $-0.17_{+0.10}^{-0.07}$ & $0.66_{-0.22}^{+0.18}$ & $0.17_{-0.03}^{+0.02}$ & $-0.92_{+0.07}^{-0.05}$ & $4.29_{-0.02}^{+0.02}$ & $3.69_{-0.02}^{+0.01}$ & $16.15_{-7.91}^{+8.09}$ & $44.97_{-6.34}^{+6.17}$ & $-0.71_{+0.02}^{-0.02}$ & $-4.81_{+0.37}^{-0.30}$ &  54914.00  \\\\
SN2009iz & $0.17_{-0.10}^{+0.15}$ & $0.08_{-0.02}^{+0.02}$ & $0.14_{-0.04}^{+0.03}$ & $1.40_{-1.55}^{+1.62}$ & $3.93_{-0.05}^{+0.04}$ & $3.74_{-0.01}^{+0.01}$ & $15.30_{-7.67}^{+8.57}$ & $43.25_{-5.24}^{+6.54}$ & $-0.63_{+0.03}^{-0.03}$ & $-8.34_{+0.84}^{-1.12}$ &  55095.43  \\\\
SN2009jf & $0.37_{-0.11}^{+0.13}$ & $0.07_{-0.02}^{+0.02}$ & $0.14_{-0.03}^{+0.03}$ & $1.32_{-1.35}^{+1.73}$ & $3.95_{-0.02}^{+0.02}$ & $3.61_{-0.01}^{+0.01}$ & $13.80_{-6.88}^{+8.42}$ & $42.05_{-4.20}^{+5.74}$ & $-0.76_{+0.04}^{-0.03}$ & $-4.53_{+0.22}^{-0.25}$ &  55103.67  \\\\
SN2010bh & $0.07_{-0.11}^{+0.14}$ & $0.12_{-0.04}^{+0.04}$ & $0.15_{-0.04}^{+0.03}$ & $-0.07_{+0.20}^{-0.21}$ & $4.87_{-0.10}^{+0.09}$ & $3.67_{-0.02}^{+0.02}$ & $8.12_{-1.12}^{+1.06}$ & $45.95_{-0.52}^{+0.51}$ & $-0.54_{+0.05}^{-0.04}$ & $-0.44_{+0.04}^{-0.02}$ &  55272.00  \\\\
SN2010et & $-0.28_{+0.08}^{-0.07}$ & $0.06_{-0.01}^{+0.01}$ & $0.17_{-0.03}^{+0.02}$ & $1.57_{-2.23}^{+1.51}$ & $3.89_{-0.02}^{+0.02}$ & $3.51_{-0.01}^{+0.01}$ & $15.68_{-7.10}^{+8.05}$ & $43.87_{-5.26}^{+6.35}$ & $-1.95_{+0.45}^{-0.53}$ & $-8.13_{+0.52}^{-0.52}$ &  55353.19  \\\\
SN2011bm & $0.59_{-0.10}^{+0.12}$ & $0.16_{-0.04}^{+0.04}$ & $0.14_{-0.03}^{+0.04}$ & $2.15_{-1.80}^{+1.26}$ & $3.91_{-0.02}^{+0.01}$ & $3.36_{-0.22}^{+0.22}$ & $8.42_{-2.33}^{+3.76}$ & $48.29_{-1.53}^{+2.69}$ & $-0.93_{+0.04}^{-0.03}$ & $-3.09_{+0.83}^{-1.75}$ &  55646.03  \\\\
SN2012P & $0.13_{-0.09}^{+0.13}$ & $0.02_{-0.01}^{+0.00}$ & $0.15_{-0.04}^{+0.03}$ & $1.40_{-1.51}^{+1.64}$ & $3.73_{-0.02}^{+0.02}$ & $3.54_{-0.00}^{+0.00}$ & $16.93_{-7.86}^{+7.82}$ & $43.77_{-5.48}^{+6.60}$ & $-0.87_{+0.04}^{-0.03}$ & $-9.51_{+0.68}^{-0.77}$ &  55939.48  \\\\
SN2012ap & $0.47_{-0.11}^{+0.12}$ & $0.05_{-0.01}^{+0.01}$ & $0.18_{-0.03}^{+0.02}$ & $-0.89_{+0.11}^{-0.10}$ & $4.36_{-0.05}^{+0.20}$ & $3.55_{-0.02}^{+0.04}$ & $15.89_{-9.42}^{+9.14}$ & $41.35_{-4.09}^{+5.93}$ & $-0.82_{+0.06}^{-0.05}$ & $-6.14_{+1.95}^{-1.43}$ &  55967.25  \\\\
PTF12gzk & $-0.25_{+0.05}^{-0.02}$ & $0.93_{-0.11}^{+0.05}$ & $0.05_{-0.00}^{+0.01}$ & $1.02_{-1.35}^{+1.67}$ & $3.98_{-0.02}^{+0.03}$ & $3.52_{-0.31}^{+0.28}$ & $18.09_{-9.45}^{+7.68}$ & $42.04_{-4.30}^{+4.43}$ & $-1.09_{+0.11}^{-0.07}$ & $-0.73_{+0.03}^{-0.04}$ &  56132.26  \\\\
iPTF13bvn & $-0.11_{+0.14}^{-0.08}$ & $0.05_{-0.01}^{+0.01}$ & $0.15_{-0.04}^{+0.03}$ & $1.64_{-1.60}^{+1.54}$ & $3.82_{-0.01}^{+0.01}$ & $3.56_{-0.01}^{+0.00}$ & $19.76_{-9.42}^{+6.56}$ & $40.38_{-3.31}^{+4.32}$ & $-0.66_{+0.02}^{-0.02}$ & $-1.92_{+0.07}^{-0.08}$ &  56459.24  \\\\
SN2013ge & $0.44_{-0.08}^{+0.12}$ & $0.04_{-0.01}^{+0.01}$ & $0.15_{-0.04}^{+0.03}$ & $1.27_{-1.44}^{+1.66}$ & $3.91_{-0.02}^{+0.02}$ & $3.54_{-0.01}^{+0.01}$ & $19.49_{-9.53}^{+6.74}$ & $41.65_{-4.05}^{+5.28}$ & $-0.62_{+0.03}^{-0.02}$ & $-11.39_{+0.49}^{-0.28}$ &  56607.00  \\\\
SN2014L & $0.53_{-0.05}^{+0.05}$ & $0.09_{-0.01}^{+0.02}$ & $0.17_{-0.02}^{+0.02}$ & $-0.88_{+0.07}^{-0.06}$ & $4.53_{-0.02}^{+0.02}$ & $3.55_{-0.01}^{+0.01}$ & $18.22_{-5.78}^{+6.89}$ & $48.30_{-7.24}^{+4.17}$ & $-0.81_{+0.03}^{-0.02}$ & $-3.33_{+0.16}^{-0.16}$ &  56684.72  \\\\
SN2014ad & $0.07_{-0.07}^{+0.07}$ & $0.16_{-0.03}^{+0.03}$ & $0.17_{-0.03}^{+0.02}$ & $-0.85_{+0.15}^{-0.08}$ & $4.30_{-0.05}^{+0.06}$ & $3.66_{-0.01}^{+0.01}$ & $15.16_{-9.29}^{+9.26}$ & $44.02_{-5.55}^{+5.95}$ & $-0.55_{+0.04}^{-0.03}$ & $-5.47_{+0.98}^{-0.77}$ &  56729.91  \\\\
SN2015ap & $0.26_{-0.11}^{+0.15}$ & $0.63_{-0.21}^{+0.19}$ & $0.10_{-0.03}^{+0.04}$ & $1.21_{-1.31}^{+1.60}$ & $4.45_{-0.03}^{+0.02}$ & $3.47_{-0.28}^{+0.20}$ & $17.34_{-8.97}^{+7.46}$ & $42.50_{-4.74}^{+6.33}$ & $-0.80_{+0.07}^{-0.05}$ & $-3.61_{+0.25}^{-0.34}$ &  57274.97  \\\\
PTF15dtg & $0.35_{-0.11}^{+0.17}$ & $0.17_{-0.05}^{+0.05}$ & $0.14_{-0.04}^{+0.04}$ & $2.11_{-1.72}^{+1.22}$ & $3.92_{-0.04}^{+0.03}$ & $3.80_{-0.20}^{+0.02}$ & $5.37_{-0.55}^{+0.52}$ & $46.53_{-0.46}^{+0.45}$ & $-0.79_{+0.04}^{-0.04}$ & $-2.27_{+0.34}^{-0.53}$ &  57333.43  \\\\
SN2016bau & $-0.83_{+0.17}^{-0.09}$ & $0.68_{-0.14}^{+0.16}$ & $0.08_{-0.02}^{+0.04}$ & $2.00_{-1.33}^{+1.24}$ & $3.32_{-0.05}^{+0.06}$ & $4.22_{-0.05}^{+0.06}$ & $17.56_{-7.42}^{+7.66}$ & $45.29_{-5.94}^{+6.10}$ & $-1.77_{+0.47}^{-0.18}$ & $-4.67_{+0.30}^{-0.27}$ &  57463.61  \\\\
SN2016blz & $0.19_{-0.34}^{+0.42}$ & $0.58_{-0.29}^{+0.24}$ & $0.11_{-0.03}^{+0.05}$ & $2.28_{-1.57}^{+1.07}$ & $4.11_{-0.23}^{+0.42}$ & $4.04_{-0.23}^{+0.29}$ & $18.79_{-9.15}^{+7.40}$ & $40.58_{-3.48}^{+6.12}$ & $-1.43_{+0.33}^{-0.37}$ & $-7.10_{+1.03}^{-1.45}$ &  57488.44  \\\\
SN2016coi & $-0.08_{+0.10}^{-0.05}$ & $0.53_{-0.16}^{+0.18}$ & $0.19_{-0.01}^{+0.01}$ & $-0.97_{+0.08}^{-0.02}$ & $4.09_{-0.03}^{+0.03}$ & $3.80_{-0.03}^{+0.02}$ & $20.42_{-8.45}^{+6.28}$ & $45.87_{-6.35}^{+5.80}$ & $-0.45_{+0.02}^{-0.02}$ & $-4.25_{+0.39}^{-0.62}$ &  57535.77  \\\\
SN2017ein & $0.04_{-0.10}^{+0.12}$ & $0.02_{-0.01}^{+0.00}$ & $0.15_{-0.04}^{+0.03}$ & $1.62_{-1.66}^{+1.52}$ & $3.77_{-0.03}^{+0.03}$ & $3.54_{-0.01}^{+0.01}$ & $17.88_{-11.50}^{+7.91}$ & $41.04_{-3.67}^{+3.32}$ & $-0.55_{+0.03}^{-0.02}$ & $-2.25_{+0.27}^{-0.28}$ &  57898.77  \\\\
SN2017fwm & $-0.86_{+0.15}^{-0.09}$ & $0.26_{-0.07}^{+0.13}$ & $0.08_{-0.02}^{+0.03}$ & $2.53_{-1.99}^{+0.95}$ & $4.83_{-0.11}^{+0.10}$ & $3.79_{-0.46}^{+0.26}$ & $22.26_{-7.18}^{+4.94}$ & $40.95_{-3.48}^{+3.30}$ & $-0.67_{+0.11}^{-0.07}$ & $-0.33_{+0.04}^{-0.09}$ &  57965.56  \\\\

\hline
\end{tabular}
\end{scriptsize}
\caption{
Best-fit parameters and 68\,\% uncertainties for the SBO+NiCo model.\label{tab:mosfit-sbonico}
}
\end{table*}
\end{turnpage}

\begin{turnpage}
\begin{table*}
\centering
\begin{scriptsize}
\begin{tabular}{l|ccccccccccc}
\hline
\hline
Source name & $\alpha$ & $\beta$ & $\log\, t_{\rm peak}$ & $\log\, \L_{\rm scale}$ & $\kappa\,({\rm cm}^{2}\,{\rm g}^{-1})$ & $\log\, \kappa_\gamma\,({\rm cm}^{2}\,{\rm g}^{-1})$ & $\log\, M_{\rm ej}\,(M_\odot)$ & $\log\, v_{\rm ej}\,({\rm km\,s}^{-1})$ & $\log\, \sigma$ & $t_{\rm exp}\,{\rm (days)}$ & $t_0$ (MJD) \\
\hline
SN2009bb & $6.42_{-1.04}^{+2.01}$ & $6.19_{-1.10}^{+1.93}$ & $1.75_{-0.11}^{+0.11}$ & $43.21_{-0.05}^{+0.07}$ & $0.14_{-0.06}^{+0.01}$ & $0.34_{-0.41}^{+0.85}$ & $-0.11_{+0.23}^{-0.04}$ & $4.33_{-0.02}^{+0.01}$ & $-0.69_{+0.02}^{-0.02}$ & $-4.84_{+0.34}^{-0.42}$ &  54914.00  \\\\
SN2009iz & $8.20_{-1.24}^{+1.02}$ & $2.30_{-0.42}^{+0.39}$ & $1.11_{-0.07}^{+0.06}$ & $43.51_{-0.18}^{+0.16}$ & $0.10_{-0.03}^{+0.04}$ & $2.65_{-1.67}^{+0.89}$ & $-0.73_{+0.17}^{-0.14}$ & $3.74_{-0.04}^{+0.04}$ & $-0.68_{+0.04}^{-0.03}$ & $-16.27_{+1.81}^{-1.74}$ &  55095.43  \\\\
SN2009jf & $4.46_{-1.11}^{+1.97}$ & $0.90_{-0.08}^{+0.10}$ & $0.70_{-0.10}^{+0.09}$ & $43.07_{-0.06}^{+0.08}$ & $0.10_{-0.03}^{+0.05}$ & $1.52_{-1.55}^{+1.43}$ & $0.10_{-0.17}^{+0.15}$ & $3.75_{-0.02}^{+0.02}$ & $-0.89_{+0.04}^{-0.03}$ & $-8.36_{+0.87}^{-1.24}$ &  55103.67  \\\\
SN2010bh & $3.73_{-0.80}^{+1.45}$ & $3.87_{-1.16}^{+0.98}$ & $-0.78_{+0.21}^{-0.22}$ & $47.74_{-0.41}^{+0.17}$ & $0.17_{-0.03}^{+0.02}$ & $-0.99_{+1.96}^{+0.09}$ & $1.45_{-0.12}^{+0.12}$ & $4.22_{-0.03}^{+0.05}$ & $-0.39_{+0.05}^{-0.04}$ & $-0.45_{+0.03}^{-0.02}$ &  55272.00  \\\\
SN2010et & $5.55_{-3.22}^{+2.90}$ & $0.64_{-0.03}^{+0.05}$ & $-0.29_{+0.31}^{-0.25}$ & $42.76_{-0.16}^{+0.19}$ & $0.14_{-0.05}^{+0.04}$ & $2.43_{-1.51}^{+1.00}$ & $-0.54_{+0.20}^{-0.12}$ & $3.92_{-0.03}^{+0.02}$ & $-1.46_{+0.20}^{-0.24}$ & $-5.08_{+0.50}^{-1.00}$ &  55353.19  \\\\
SN2011bm & $4.67_{-1.02}^{+1.22}$ & $5.12_{-1.01}^{+1.04}$ & $1.44_{-0.17}^{+0.23}$ & $44.01_{-0.57}^{+0.47}$ & $0.09_{-0.02}^{+0.03}$ & $0.14_{-0.69}^{+1.27}$ & $1.22_{-0.15}^{+0.15}$ & $3.71_{-0.02}^{+0.02}$ & $-0.84_{+0.03}^{-0.03}$ & $-4.26_{+0.77}^{-1.08}$ &  55646.03  \\\\
SN2012P & $6.87_{-1.39}^{+1.67}$ & $1.08_{-0.12}^{+0.11}$ & $0.88_{-0.05}^{+0.04}$ & $42.34_{-0.07}^{+0.07}$ & $0.11_{-0.03}^{+0.03}$ & $2.13_{-1.64}^{+1.19}$ & $-0.82_{+0.16}^{-0.11}$ & $3.69_{-0.02}^{+0.02}$ & $-0.91_{+0.04}^{-0.03}$ & $-12.61_{+1.21}^{-1.21}$ &  55939.48  \\\\
SN2012ap & $7.90_{-1.65}^{+1.28}$ & $6.58_{-1.43}^{+1.20}$ & $1.63_{-0.10}^{+0.10}$ & $43.65_{-0.25}^{+0.20}$ & $0.14_{-0.05}^{+0.04}$ & $1.73_{-1.54}^{+1.41}$ & $-0.04_{+0.18}^{-0.11}$ & $4.36_{-0.04}^{+0.04}$ & $-0.90_{+0.06}^{-0.04}$ & $-8.04_{+0.58}^{-0.45}$ &  55967.25  \\\\
PTF12gzk & $8.64_{-0.90}^{+0.74}$ & $4.70_{-0.68}^{+1.16}$ & $0.18_{-0.09}^{+0.13}$ & $47.03_{-0.31}^{+0.29}$ & $0.17_{-0.02}^{+0.02}$ & $2.64_{-0.92}^{+0.70}$ & $0.54_{-0.20}^{+0.23}$ & $3.67_{-0.04}^{+0.05}$ & $-1.45_{+0.14}^{-0.10}$ & $-0.71_{+0.19}^{-0.15}$ &  56132.26  \\\\
iPTF13bvn & $9.49_{-0.64}^{+0.35}$ & $7.24_{-0.66}^{+0.53}$ & $0.89_{-0.06}^{+0.05}$ & $45.07_{-0.14}^{+0.12}$ & $0.12_{-0.04}^{+0.04}$ & $1.61_{-1.55}^{+1.49}$ & $0.80_{-0.13}^{+0.19}$ & $3.54_{-0.01}^{+0.01}$ & $-0.56_{+0.02}^{-0.01}$ & $-2.34_{+0.25}^{-0.28}$ &  56459.24  \\\\
SN2013ge & $3.11_{-0.86}^{+1.09}$ & $3.80_{-0.77}^{+0.84}$ & $0.90_{-0.14}^{+0.10}$ & $44.27_{-0.30}^{+0.34}$ & $0.17_{-0.02}^{+0.02}$ & $-0.30_{-0.58}^{+1.35}$ & $1.27_{-0.11}^{+0.10}$ & $3.80_{-0.04}^{+0.03}$ & $-0.57_{+0.03}^{-0.04}$ & $-4.49_{+0.99}^{-1.34}$ &  56607.00  \\\\
SN2014L & $4.42_{-3.01}^{+3.30}$ & $0.20_{-0.06}^{+0.06}$ & $-1.08_{+0.26}^{-0.32}$ & $43.18_{-0.14}^{+0.17}$ & $0.16_{-0.03}^{+0.03}$ & $-0.87_{+0.08}^{-0.08}$ & $0.20_{-0.07}^{+0.11}$ & $4.53_{-0.01}^{+0.02}$ & $-0.84_{+0.03}^{-0.02}$ & $-3.09_{+0.23}^{-0.23}$ &  56684.72  \\\\
SN2014ad & $2.47_{-0.71}^{+1.86}$ & $1.97_{-0.56}^{+1.53}$ & $1.22_{-0.24}^{+0.26}$ & $43.11_{-0.13}^{+0.19}$ & $0.11_{-0.03}^{+0.04}$ & $2.43_{-1.64}^{+0.93}$ & $-0.25_{+0.14}^{-0.14}$ & $4.30_{-0.09}^{+0.05}$ & $-0.49_{+0.04}^{-0.04}$ & $-5.33_{+1.09}^{-2.81}$ &  56729.91  \\\\
SN2015ap & $3.63_{-0.78}^{+1.15}$ & $2.99_{-1.00}^{+1.40}$ & $1.29_{-0.29}^{+0.42}$ & $43.53_{-0.32}^{+0.23}$ & $0.13_{-0.02}^{+0.03}$ & $3.38_{-0.64}^{+0.39}$ & $0.04_{-0.08}^{+0.17}$ & $4.41_{-0.06}^{+0.04}$ & $-0.74_{+0.06}^{-0.07}$ & $-4.53_{+0.91}^{-1.31}$ &  57274.97  \\\\
PTF15dtg & $4.73_{-1.34}^{+1.30}$ & $7.54_{-1.34}^{+1.25}$ & $-1.29_{+0.43}^{-0.39}$ & $40.65_{-1.51}^{+1.65}$ & $0.17_{-0.03}^{+0.02}$ & $2.96_{-0.79}^{+0.63}$ & $0.63_{-0.08}^{+0.09}$ & $3.57_{-0.05}^{+0.04}$ & $-0.65_{+0.05}^{-0.04}$ & $-7.57_{+0.98}^{-1.21}$ &  57333.43  \\\\
SN2016bau & $8.63_{-1.46}^{+0.84}$ & $7.61_{-1.62}^{+1.19}$ & $1.68_{-0.34}^{+0.19}$ & $46.89_{-0.66}^{+0.63}$ & $0.16_{-0.02}^{+0.02}$ & $1.86_{-1.01}^{+0.89}$ & $0.36_{-0.30}^{+0.18}$ & $4.75_{-0.19}^{+0.16}$ & $-0.98_{+0.21}^{-0.70}$ & $-6.43_{+1.69}^{-2.70}$ &  57463.61  \\\\
SN2016blz & $5.64_{-3.85}^{+3.13}$ & $7.31_{-1.44}^{+1.34}$ & $-0.43_{-0.26}^{+2.14}$ & $45.44_{-1.94}^{+1.35}$ & $0.14_{-0.04}^{+0.03}$ & $0.82_{-0.97}^{+2.00}$ & $0.67_{-1.27}^{+0.71}$ & $3.54_{-0.25}^{+0.35}$ & $-0.78_{+0.21}^{-0.16}$ & $-7.36_{+3.26}^{-3.22}$ &  57488.44  \\\\
SN2016coi & $3.53_{-1.41}^{+1.20}$ & $2.00_{-0.91}^{+0.81}$ & $1.30_{-0.10}^{+0.09}$ & $43.07_{-0.25}^{+0.25}$ & $0.11_{-0.03}^{+0.04}$ & $-0.66_{+1.44}^{-0.14}$ & $-0.40_{+0.16}^{-0.13}$ & $4.04_{-0.03}^{+0.03}$ & $-0.46_{+0.02}^{-0.02}$ & $-7.23_{+1.11}^{-1.34}$ &  57535.77  \\\\
SN2017ein & $9.01_{-1.17}^{+0.66}$ & $7.64_{-0.92}^{+0.71}$ & $1.03_{-0.07}^{+0.08}$ & $44.49_{-0.31}^{+0.23}$ & $0.15_{-0.05}^{+0.03}$ & $2.17_{-1.74}^{+1.20}$ & $0.90_{-0.13}^{+0.16}$ & $3.47_{-0.03}^{+0.04}$ & $-0.47_{+0.03}^{-0.02}$ & $-2.87_{+0.58}^{-0.46}$ &  57898.77  \\\\
SN2017fwm & $6.16_{-3.36}^{+2.48}$ & $4.63_{-1.52}^{+1.81}$ & $-0.38_{+0.24}^{-0.27}$ & $46.10_{-1.18}^{+1.05}$ & $0.13_{-0.05}^{+0.04}$ & $2.01_{-1.55}^{+1.24}$ & $-0.56_{-0.27}^{+1.54}$ & $4.19_{-0.14}^{+0.18}$ & $-0.77_{+0.15}^{-0.08}$ & $-0.19_{+0.07}^{-0.24}$ &  57965.56  \\\\
\hline
\end{tabular}
\end{scriptsize}
\caption{Best-fit parameters and 68\,\% uncertainties for the exppow model. \label{tab:mosfit-exppow}}
\end{table*}
\end{turnpage}

\subsection{Comparison to the literature}
Here, we compare the fit parameters for time of explosion,  $t_\mathrm{exp}$, ejecta mass, $M_\mathrm{ej}$, and nickel mass $M_\mathrm{Ni}$ that we obtain with \mosfit with those found in the literature (see Tab.~1 of the main text for the relevant references).\footnote{For SN2015ap, we use the results obtained in Ref.~\cite{2019MNRAS.485.1559P}.}
Currently, there are no publications where these details are examined for SN2009iz, SN2016bau, SN2016blz, and SN2017fwm.
For the other SNe, the explosion times obtained with \mosfit are generally in agreement with those from the literature, 
at least for one of the three chosen models.
The largest discrepancies are found for two SNe, SN2013ge and PTFdtg.
For SN2013ge, the NiCo and SBO+NiCo models provide much earlier explosion times than the exppow model or the literature value. 
The reason is that the first two models predict a gradual flux decline towards earlier times that does not violate the upper limit.
The exppow model, in the other hand, 
shows a very steep rise and places the explosion between the 
first data point and the upper limit (as done in the literature).
For PTF15dtg, a clear bump is seen for the first data points, 
which is well modelled with the SBO component, but missed by the NiCo and exppow models.
This is also reflected in the lower best-fit value of the $\sigma$ paramter, i.e., a lower value of additional uncertainty is required in order for the fit to reach a good fit quality. 
The exppow and NiCo also predict much earlier explosion times than the SBO+NiCo model, which agrees better with the literature estimate. 
Futhermore, the lower $\sigma$ value for the SBO+NiCo model is also observed for SN2010bh, where the light curve also shows a characteristic bump feature. 

Regarding the other two explosion parameters, $M_\mathrm{ej}$ and $M_\mathrm{Ni}$, we find generally good agreement for all sources when the uncertainties are taken into account, despite that the results from different authors depend on a range of assumptions in the modeling approaches, so values from different studies cannot be easily compared directly. 

\begin{figure*}
    \centering
    \includegraphics[width = .9\linewidth]{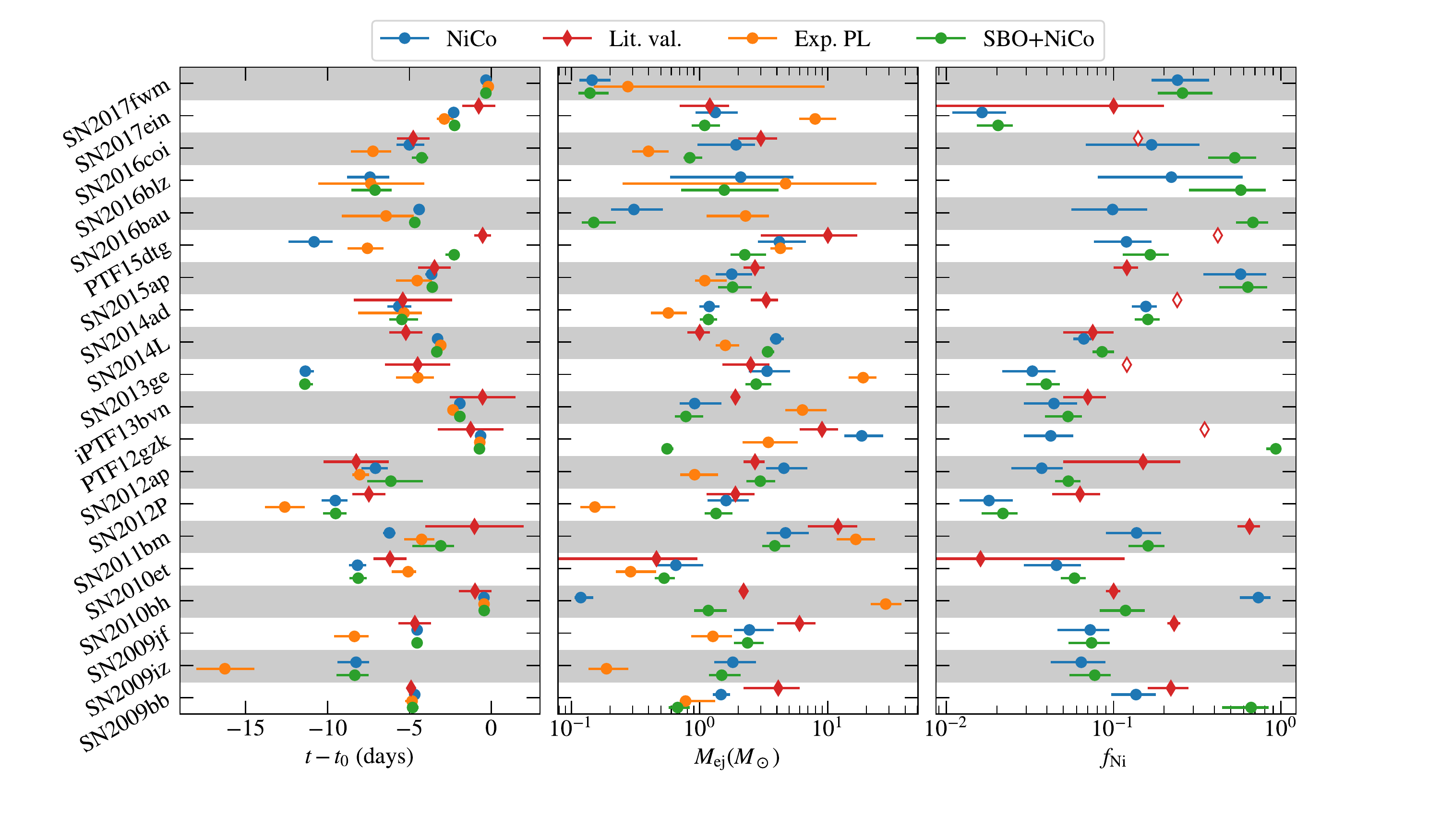}
    \caption{Comparison between \mosfit results and literature values. Open symbols indicate that the values are given without an estimate of the uncertainty. }
    \label{fig:comp_literature}
\end{figure*}{}

\section{$\gamma$-ray Results for individual SNe}

\begin{figure*}
    \centering
    \includegraphics[width=.9\linewidth]{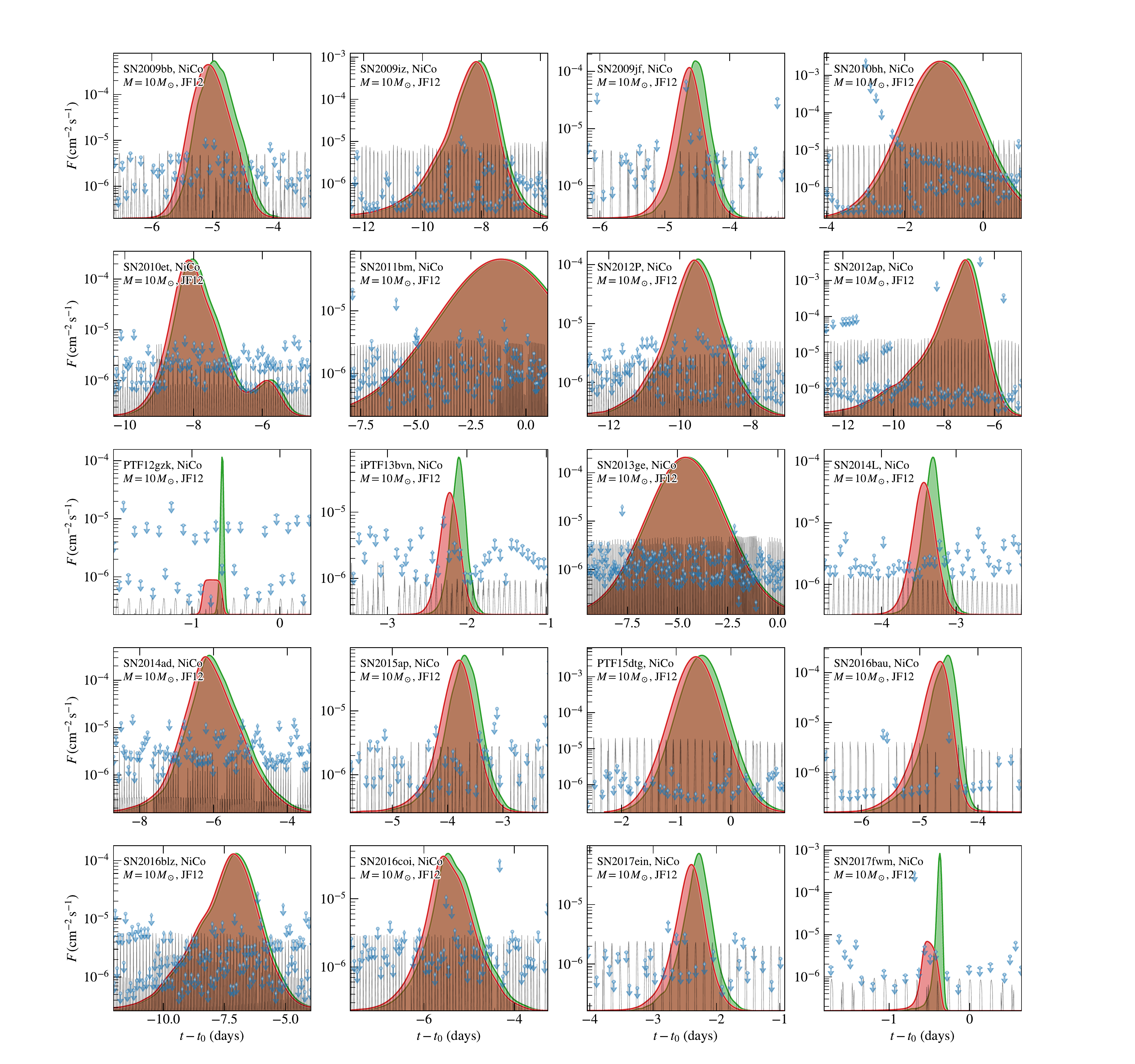}
    \caption{\textit{Fermi}-LAT light curves for the individual SNe. The upper limits on the flux in the first energy bin (60-83\,MeV) of the analysis are shown as blue arrows. The relative exposure of the LAT (normalized to the maximum exposure within the 60~day analysis interval) is shown with grey lines. 
    It can vary between orbits due to the alternating rocking angle of the satellite towards the northern and southern celestial poles.
    Observational gaps are due to, e.g., pointed operations of the satellite during target of opportunity observations or automated repointing for ordinary \gr bursts.
    The marginalized posterior for the explosion time extracted from the \mosfit results, $P(t_\mathrm{exp})$, is shown as a green filled line.
    Accounting for the delay between core collapse and explosion time yields the convolved posterior $P(t_\mathrm{CC})$, shown in red. 
    For the SN2010bh, SN2011b, SN2013ge, and PFT15dtg we use literature values and assume a Gaussian distribution for the posterior of $t_\mathrm{exp}$.
    }
    \label{fig:all-gr-lc}
\end{figure*}

In this section, we present result plots for all SNe in the sample. 
In Fig.~\ref{fig:all-gr-lc}, the \gr light curves of the first energy bin of the analysis are shown together with the posterior distributions for the explosion and core-collapse time from the \mosfit NiCo model (green and red filled lines, respectively). 
No time bin in the light curves shows evidence for a detection of an SN signal defined as $\mathrm{TS}\geqslant9$ in the first energy bin shown here. 
The plot further demonstrates that the number of orbits in which the potential signal might have occurred decreases significantly depending on the width of $P(t_\mathrm{cc})$.

\begin{figure*}
    \centering
    \includegraphics[width=.9\linewidth]{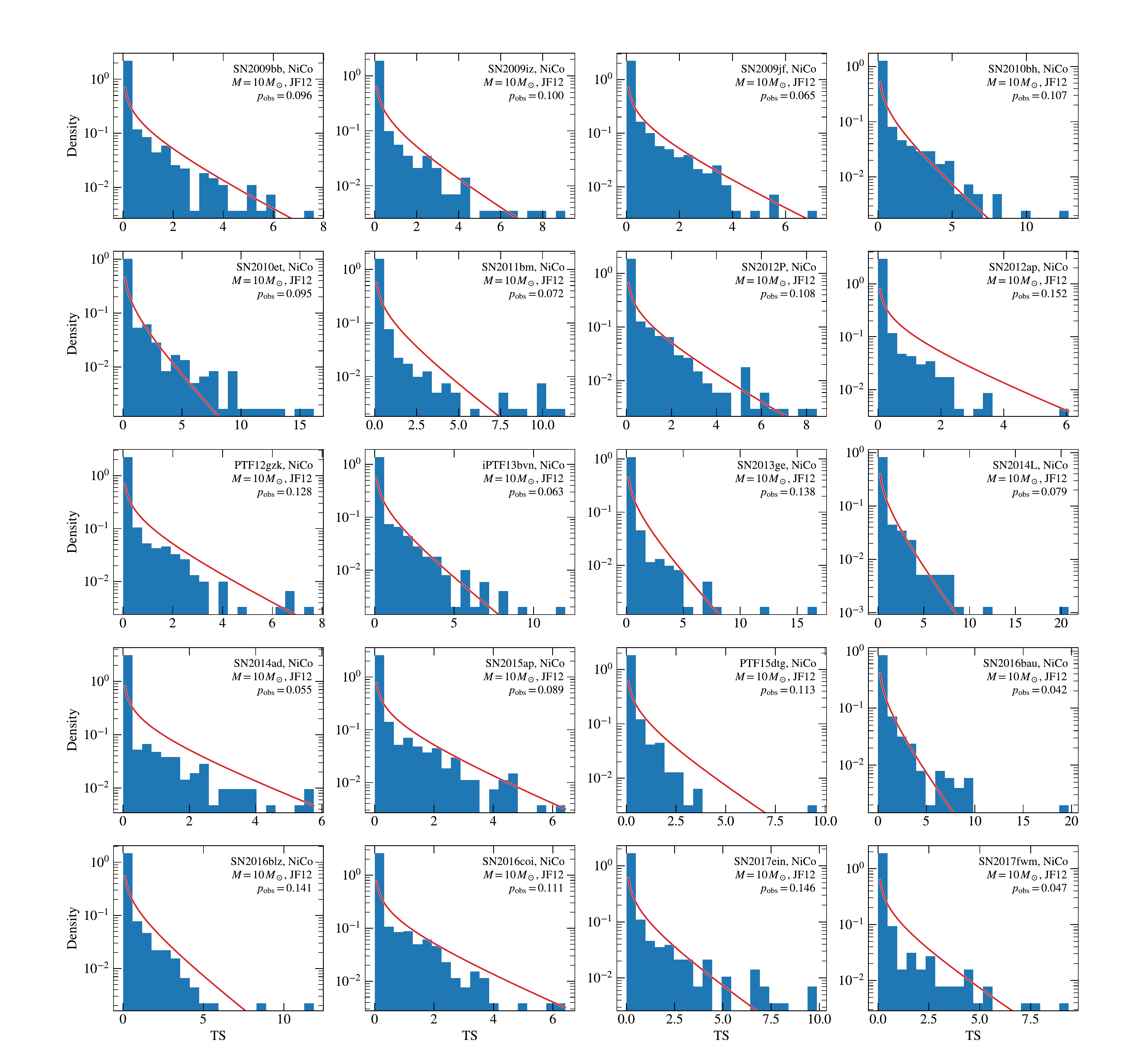}
    \caption{Null distributions for the single SNe derived from LAT orbits outside the 99\,\% quantile of $P(t_\mathrm{CC})$ for the NiCo model. The histograms show the observed $\mathrm{TS}$ values and the red solid line is the expectation from a $\chi^2_{\nu = 1} / 2$ distribution with $\nu = 1$ degree of freedom.}
    \label{fig:all-ts}
\end{figure*}

In Fig.~\ref{fig:all-ts} we show the $\mathrm{TS}$ distributions for all SNe derived from our ``Off'' data, i.e., for orbits that fall outside the 99\,\% quantile of $P(t_\mathrm{CC})$.
Therefore, these orbits should not contain a potential SN signal.
Depending on the width of $P(t_\mathrm{CC})$, the sample of ``Off'' observations contains several hundred orbits within the analyzed 60\,day interval.  
Since an SN source is modeled with one additional parameter, the normalization of the SN spectral template, we expect this null distribution to follow a $\chi^2 / 2$ distribution with 1 degree of freedom. 
Generally, the $\mathrm{TS}$ distributions follow the expected trend. 
For some SNe a tail towards higher $\mathrm{TS}$ values is observed (e.g., SN2010et) which could be 
due to unaccounted sub-threshold point sources. 
Other cases show a stronger peak towards $\mathrm{TS} = 0$ (e.g., SN2012ap) which could be due to statistical fluctuations of the data. 

\begin{figure*}
    \centering
    \includegraphics[width=.9\linewidth]{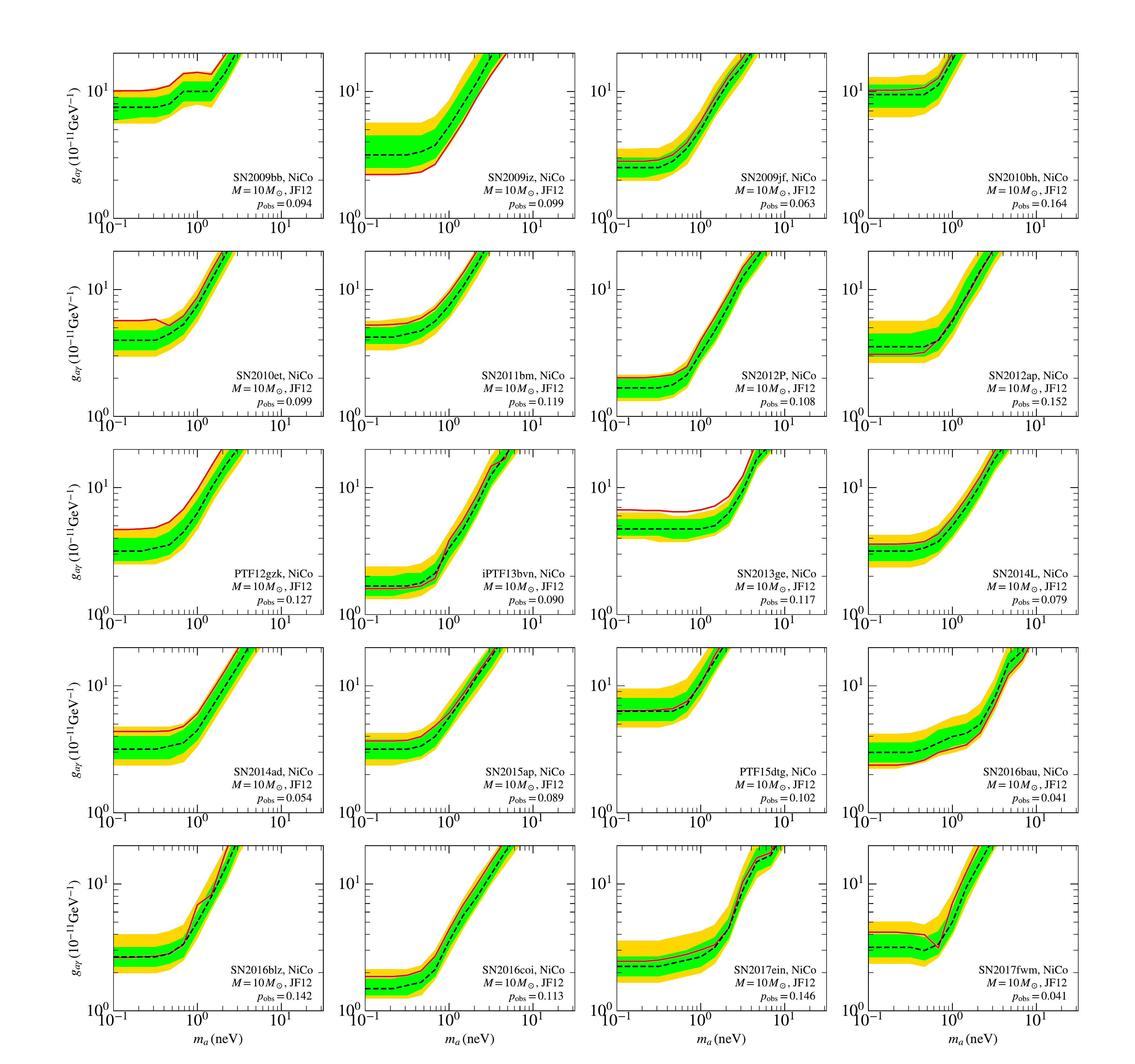}
    \caption{Expected and observed 95\,\% confidence upper limits on the photon-ALP coupling as a function of the ALP mass. The expected limits are derived from the ``Off'' orbits (see text). 
    The median expected limit is shown as a black dashed line and the 68\,\% and 95\,\% spread of the expectation as green and yellow shaded bands, respectively. 
    The observed limits are shown in red.
    The probability that the limits are valid, i.e., that the SN was in the field of view of the LAT at the time of the core collapse, is given in each panel as the $p_\mathrm{obs}$ value.
    }
    \label{fig:all-gr-bands}
\end{figure*}

We use the same ``Off'' orbits to derive the expected constraints on the photon-ALP coupling for each SN, shown as a black dashed line for the median and green and yellow bands for the 68\,\% and 95\,\% confidence bands in Fig.~\ref{fig:all-gr-bands}.
These can be compared to the individual observed upper limits from each SN, shown as red lines in Fig.~\ref{fig:all-gr-bands}.
The limits in general agree extremely well with the expectations. 
The only exception is SN2017fwm for some particular ALP masses. 
However, also on this case, the $\mathrm{TS}$ values are still too low to claim any evidence for an ALP signal. 

\section{Coverage Test}
We have tested our statistical procedure by means of a coverage test since the choice of selecting the GTI with maximum $\mathrm{TS}_{ij}$ is not unique. 
We could have selected instead a GTI where $P(t_\mathrm{CC})$ is maximal or we might have chosen to treat $t_j$ as an additional nuisance parameter. 
In the first case, however, we would disregard the uncertainty on the core-collapse time and bias ourselves against a potential detection. 
The second case is also problematic since each GTI is an independent experiment and by profiling over $t_j$ we would effectively use likelihood values from different GTIs (and therefore different data) depending on the tested value of $g_{a\gamma}$. 
From the definition of the Poisson likelihood it is obvious that this would bias the unconditionally maximized likelihood $\ln\mathcal{L}_{ijk}(\hat{g}_{a\gamma})$ towards GTIs with a low number of counts and low values of $g_{a\gamma}$.

For the coverage test,
we extract likelihood profiles $\mathcal{L}_{ijk}$ for each source $i$, GTI $j$, and energy bin $k$ as outlined in the main paper and use these profiles to compute the detection significance $\mathrm{TS}$ of a potential signal and the log-likelihood ratio test $\lambda$ in order to derive upper limits on the photon-ALP coupling. 
This procedure is closely related to what has been done in previous searches with the \fermiLAT for ALPs and dark matter in general~\cite{2016PhRvL.116p1101A,2015PhRvL.115w1301A}. 
Using the likelihood profiles instead of conducting a standard \fermiLAT analysis using the official \fermiLAT \textit{science tools} has the advantage that different ALP spectra can be tested easily; the likelihood of the model can simply be read of the $\mathcal{L}_{ijk}$ profiles for a given predicted flux in each energy bin.
It has been shown that the results are insensitive to the assumed spectrum with which the likelihood profiles are extracted in the first place~\cite{2015PhRvL.115w1301A}.
For the standard \textit{science tools}, on the other hand, one would have to repeat the entire analysis, in particular the calculation of the light curve, for each assumed model. 
In particular, this would mean that the light curves for $\sim 1000$ GTIs for each SN would have to be extracted for every considered model in our test of systematic uncertainties, described in the main part of the paper and in the next section. 
For this reason, we opt to use the likelihood-profile instead.

The goal of the coverage test is to verify that the chosen threshold for $\lambda$ provides the right level of confidence for the upper limits. 
To this end, we select two SNe (SN2012ap and SN2017fwm) with broad and narrow posterior $P(t_\mathrm{CC})$ and simulate observations with an injected \gr burst signal. 
The procedure is as follows: 
we draw random times from $P(t_\mathrm{CC})$ and select those that fall within a GTI. 
We repeat this until we have of the order of 100 randomly drawn GTIs. 
For each of these, we simulate the ROI, starting from the best-fit ROI model derived during the light curve calculation of the real data. 
On top of all background sources, we inject a ALP-induced \gr burst with a flux averaged over the GTI duration. 
The signal is injected for different values of $g_{a\gamma} = (0.1, 0.2, \dots, 1, 2)\times10^{-11}\,\mathrm{GeV}^{-1}$ for $m_a = 0.1\,$neV, a $10\,M_\odot$ progenitor, and the JF12 Galactic mangetic field model. 
For each $g_{a\gamma}$ value, GTI, and SN we repeat our analysis in the same way as with real data. 
We can now test whether our assumption is correct that the log-likelihood ratio $\lambda$ follows a $\chi^2$ distribution with 1 degree of freedom. 
We should recover our injected signal at a fraction $\alpha$ inside a $g_{a\gamma}$ range defined by the   $\Delta\lambda$ value corresponding to $\alpha$.
For example, a one-sided $95\,\%$ confidence limit corresponds to $\Delta\lambda=2.71$.

\begin{figure*}
    \centering
    \includegraphics[width=.49\linewidth]{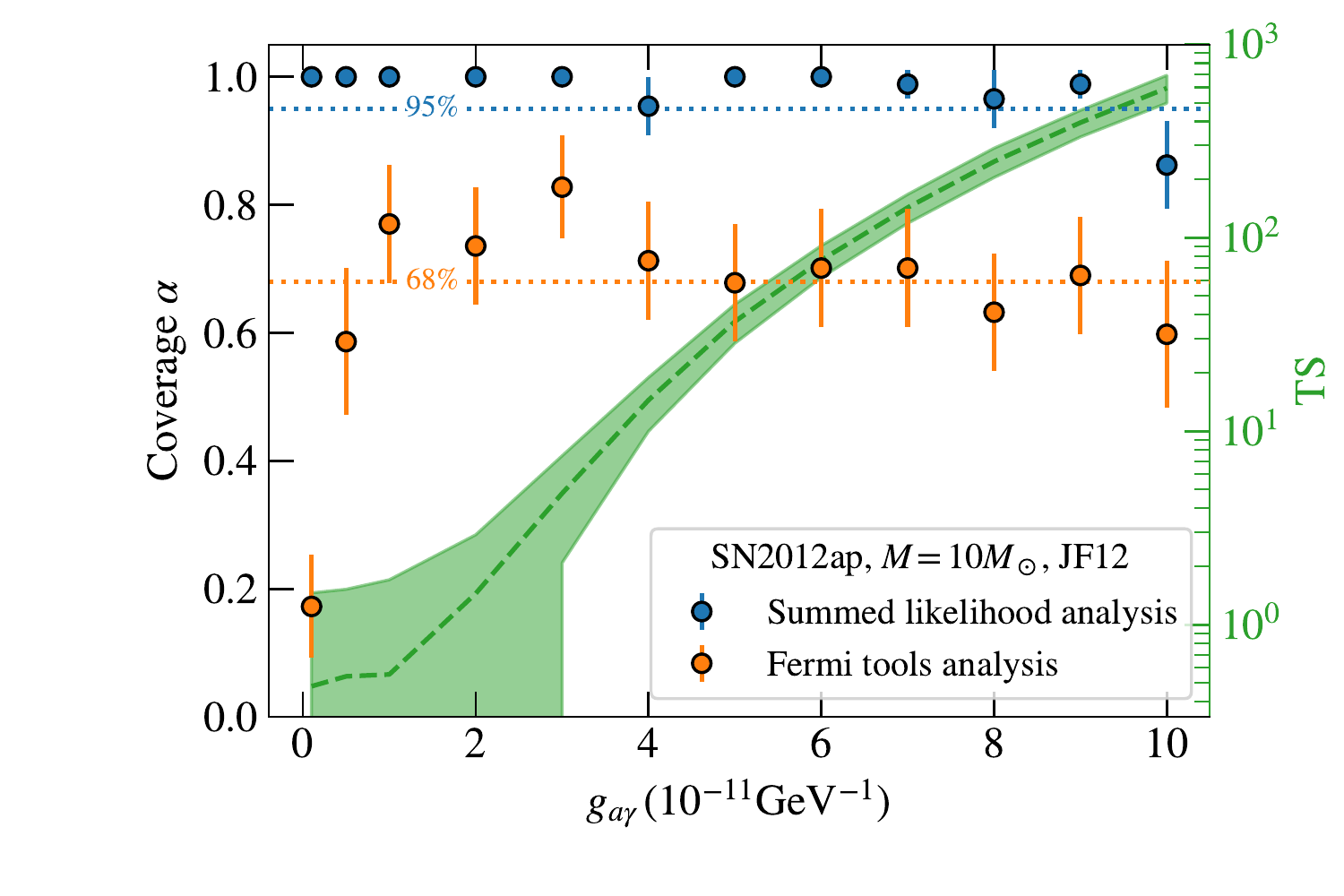}
    \includegraphics[width=.49\linewidth]{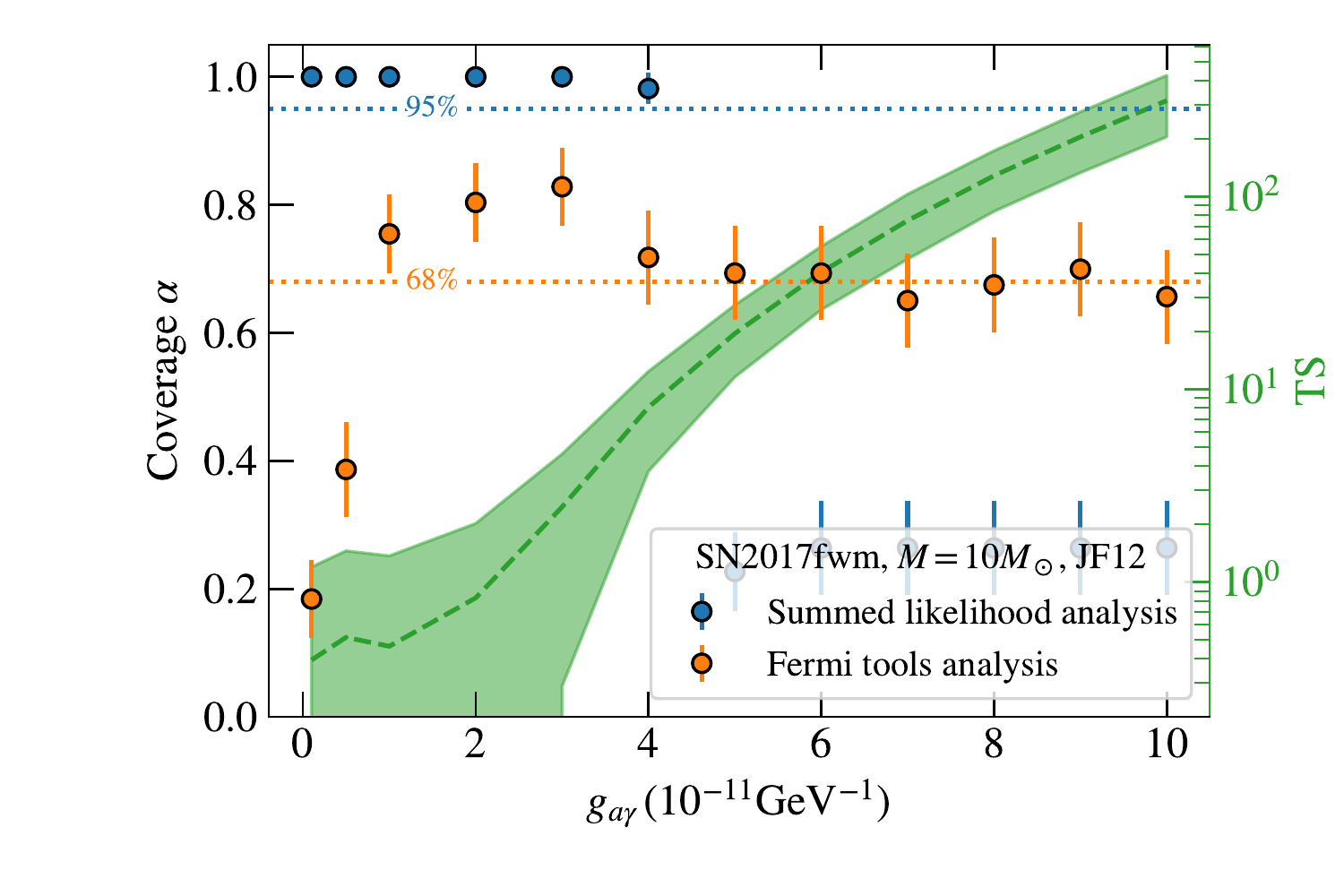}
    \caption{
    Results of the coverage test for SN2012ap (left) and SN2017fwm (right). 
    The fraction $\alpha$ with which the injected signal is included in the 95\,\% confidence interval for the summed bin-by-bin likelihood analysis is shown with blue markers. For the correct coverage, the points (within error bars) should fall on the dashed blue line, which marks a coverage of 95\,\%. As discussed in the text, we find under-coverage (points fall below the dashed blue line) once the signal is significantly detected.
    The coverage for a 68\,\% confidence interval for the standard \fermiLAT \textit{science tools} analysis is shown with orange markers. 
    The points fall on the dashed orange line at high coupling values, indicating correct coverage. 
    The detection significance of the injected signal is given by the $\mathrm{TS}$ values, calculated from the bin-by-bin likelihood curves, which are shown in green and on the second $y$ axis.
    The dashed green line represents the median of all pseudo experiments and the spread is the RMS.
    }
    \label{fig:cov-test}
\end{figure*}

\begin{figure*}
    \includegraphics[width=.9\linewidth]{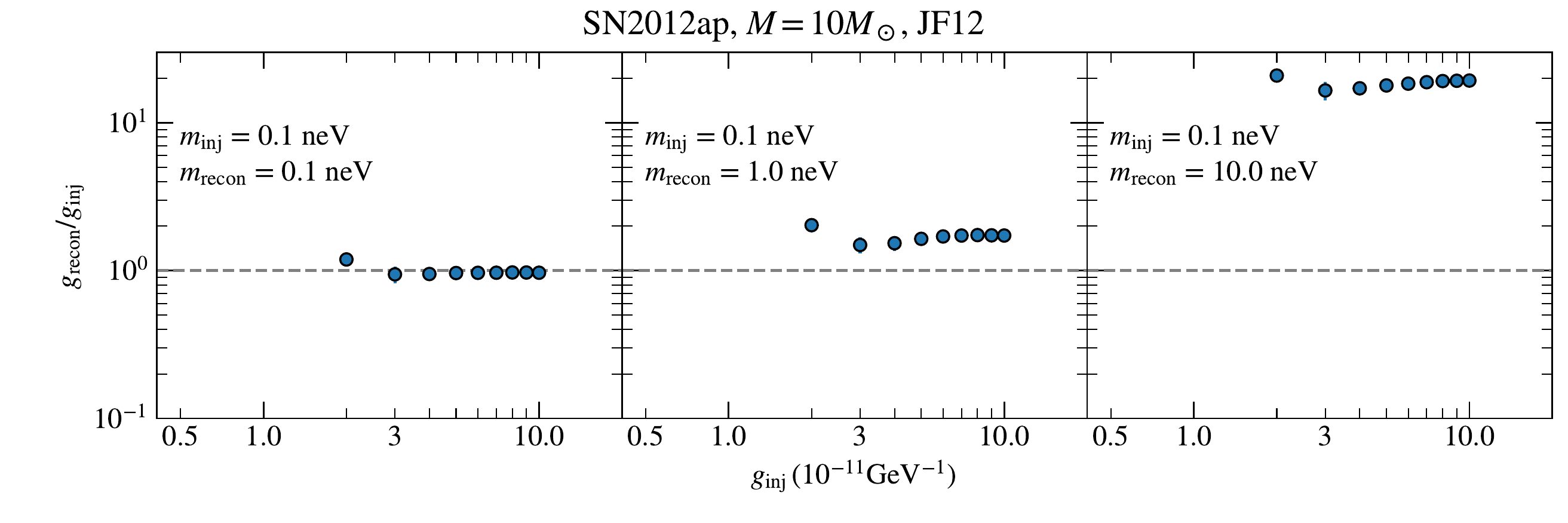}
    \includegraphics[width=.9\linewidth]{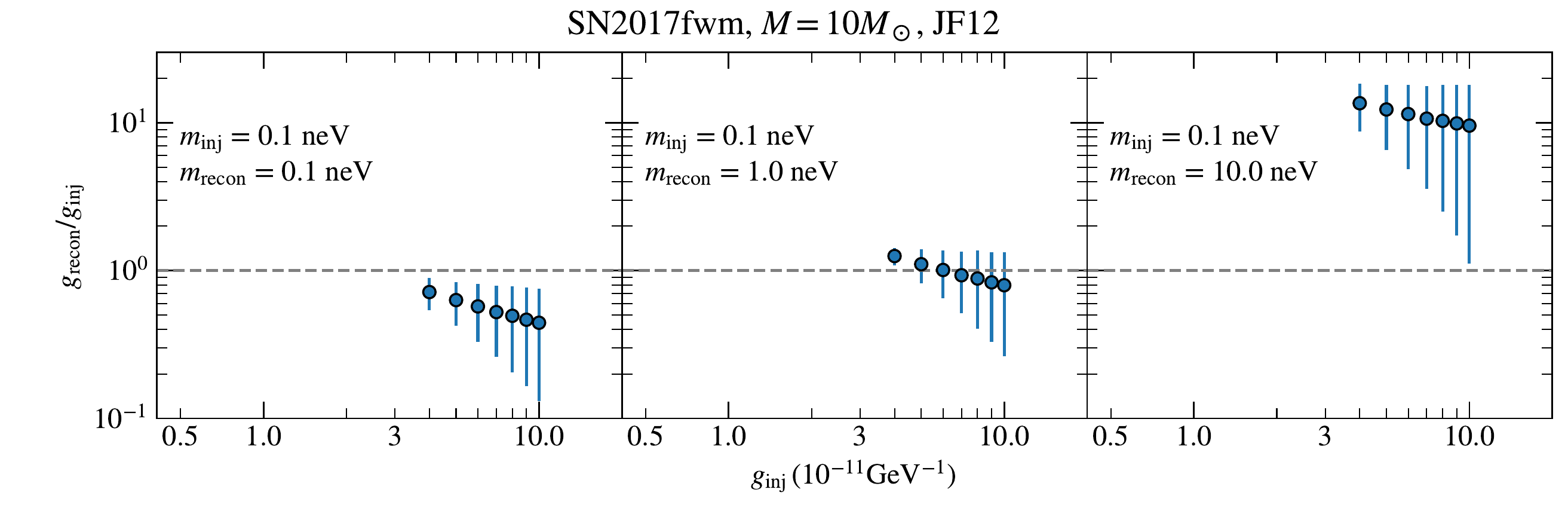}
    \caption{Reconstructed coupling values for different assumed masses in the summed likelihood analysis for SN2012ap (top) and SN2017fwm (bottom). 
    For $m_a = 0.1\,$neV (at which the signal was injected) the analysis shows a bias towards lower couplings due to the chosen interpolation scheme of the log-likelihood curves. 
    As one increases the mass, the reconstructed couplings are also higher to compensate for the reduced photon-ALP conversion in the Galactic magnetic field.}
    \label{fig:recon-sig}
\end{figure*}

We define coverage as $\alpha = N_\mathrm{in} / N_\mathrm{sim}$, where $N_{in}$ is the number of simulations where the injected signal is within the derived confidence interval (or equivalently not ruled out by the derived upper limit) and $N_\mathrm{sim}$ is the total number of simulations. 
The coverage is shown in Fig.~\ref{fig:cov-test} for both tested SNe. 
For our summed likelihood analysis that uses the bin-by-bin likelihood curves $\mathcal{L}_{ijk}$, we find the correct coverage for low values of $g_{a\gamma}$, where the signal is not significantly detected.
This is shown by the $\mathrm{TS}$ values computed from the likeihood profiles, which are plotted on the right $y$ axis. 
For couplings $g_{a\gamma} > 0.3\times10^{-11}\,\mathrm{GeV}^{-1}$, the signal is significantly detected, $\mathrm{TS} \gtrsim 25$, and the bin-by-bin analysis results in under-coverage, i.e., 
the injected signal is included in the derived confidence interval at a fraction $< 95\,\%$ (corresponding to $\Delta\lambda = 2.71$ for upper limits and $\Delta\lambda = 3.84$ for a two-sided confidence interval). 
This becomes more severe as the coupling is further increased. 
The reason is of computational nature: we use quadratic splines to interpolate the $\ln\mathcal{L}_{ijk}$ curves but for larger couplings, where the signal is significantly detected, the log-likelihood curves are given by very narrow parabolas. 
Small mis-representations of the parabolas by the interpolated curves have therefore a large effect on the derived confidence intervals. 
To further illustrate this, we compare the reconstructed coupling values with the injected ones in  Fig.~\ref{fig:recon-sig}. 
When assuming the same mass for the reconstruction as the one of the injected signal, we observe that for high values of $g_{a\gamma}$, the reconstructed values are lower than the injected ones but still close in an absolute sense.
We also observe the trend that for higher assumed masses in reconstruction the reconstructed coupling moves to higher values in order to compensate the lower photon-ALP oscillation probability. 

Lastly, we perform a test of the standard \fermiLAT science tools pipeline to reconstruct the signal. 
The coverage properties of this analysis are also shown in the upper panels of Fig.~\ref{fig:cov-test}, this time for a two-sided $68\,\%$ confidence interval. 
As long as the signal is not significantly detected, we find again under-coverage and the injected signals are not reconstructed reliably. 
However, as soon as the signal is strong enough, the science tools recover the injected signal at the correct rate. 

To summarize, our coverage confirms that the $\lambda$ values follow a $\chi^2$ distribution with 1 degree of freedom. 
The bin-by-bin likelihood analysis gives the correct coverage when no signal is significantly detected. 
For stronger signals, the analysis is biased towards lower values of the coupling compared to the injected ones owing to the numerical procedure chosen. 
However, the right coupling values are recovered in this case by the standard \fermiLAT science tools analysis software.

\section{Model uncertainties}

\begin{figure}
    \centering
    \includegraphics[width = .8\linewidth]{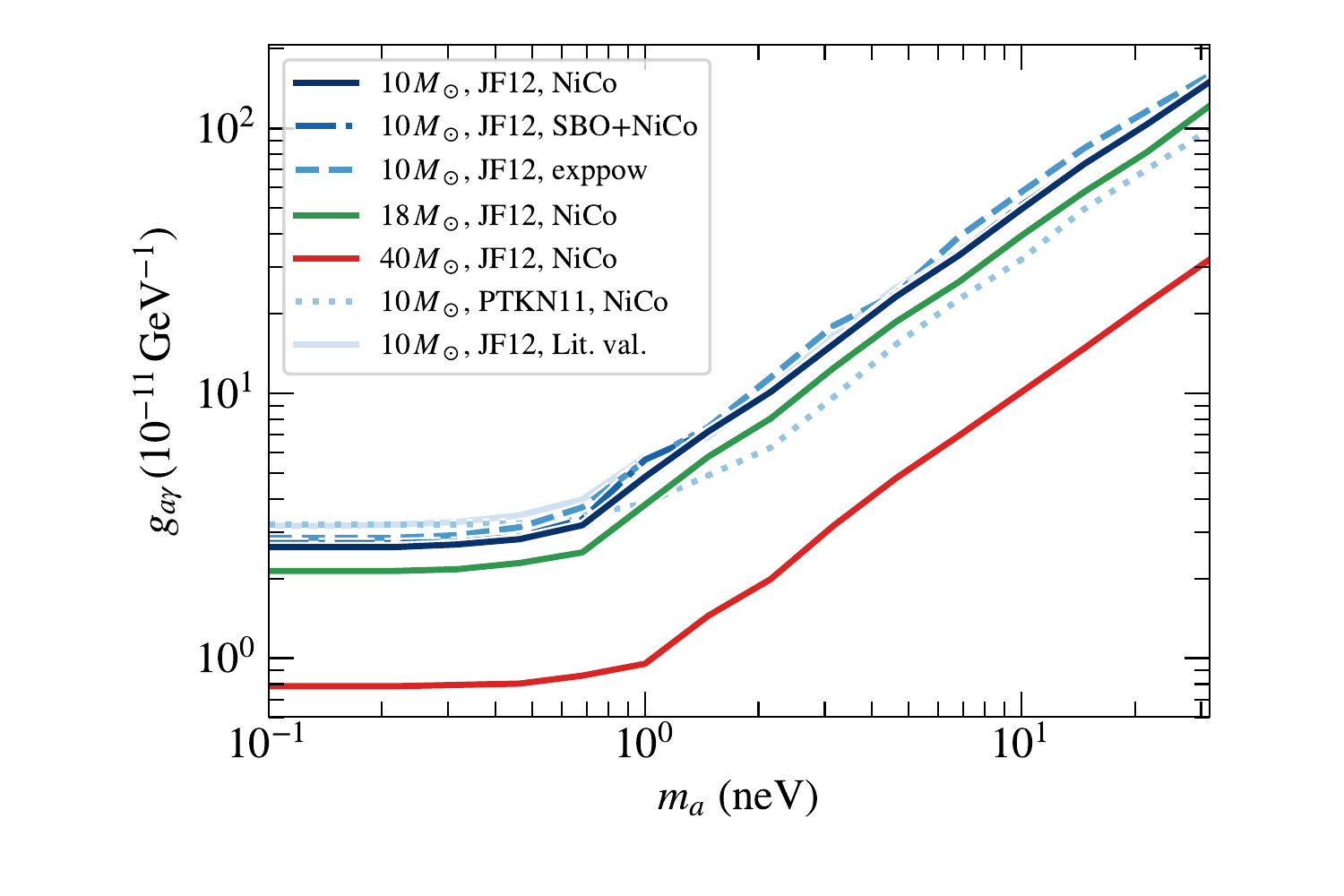}
    \caption{Median limits derived from the likelihood stacking for different choices of the Galactic magnetic field, progenitor mass, and SN engine (parameters above the lines are excluded). The probability to observe at least one SN within a given model is reported in Tab.~\ref{tab:pobsTS}.}
    \label{fig:sys-unc}
\end{figure}

In this section, we provide further details on the systematic uncertainties related to model uncertainties. 
As discussed in the main part of the paper, we identify the main unknowns in the analysis as 1) the progenitor mass for simplicity chosen to be the same for all SNe, 2) the model of the Galactic magnetic field, and 3) the model of the evolution of the optical emission of the SN after the explosion. 
To address these uncertainties, we vary our model assumptions and repeat the analysis. 
The results are shown in Fig.~\ref{fig:sys-unc} where we report the median limits of our likelihood stacking procedure. 
In terms of the progenitor mass, detailed simulations for the SN explosion and the ALP production rate have been carried out for progenitor masses of $10\,M_\odot$ and $18\,M_\odot$ in previous works~\cite{2015JCAP...02..006P,2017PhRvL.118a1103M}, which are also used here. 
We have also added preliminary results for a $40\,M_\odot$ progenitor, derived for the first 13\,s after the core collapse~\cite{40mprogenitor}, motivated by our choice of type~Ib/c SNe which are thought to be triggered by massive blue supergiants or Wolf-Rayet stars. 
In the simulations of these high-mass stars, the explosions are driven by first order hadron-quark phase transition, which is however, highly speculative. 
Furthermore, the evolution of such stars in particular in their final phases before the core collapse is also highly uncertain. The stars are likely to experience large outbursts while entering the luminous blue variable phase which is very challenging to include self-consistently in the simulations. 
Therefore, the results presented here should be regarded only as an indication how the ALP production rate could change with higher progenitor mass and further details of the full calculation will be presented elsewhere.

\begin{figure*}
    \centering
    \includegraphics[width = .9\linewidth]{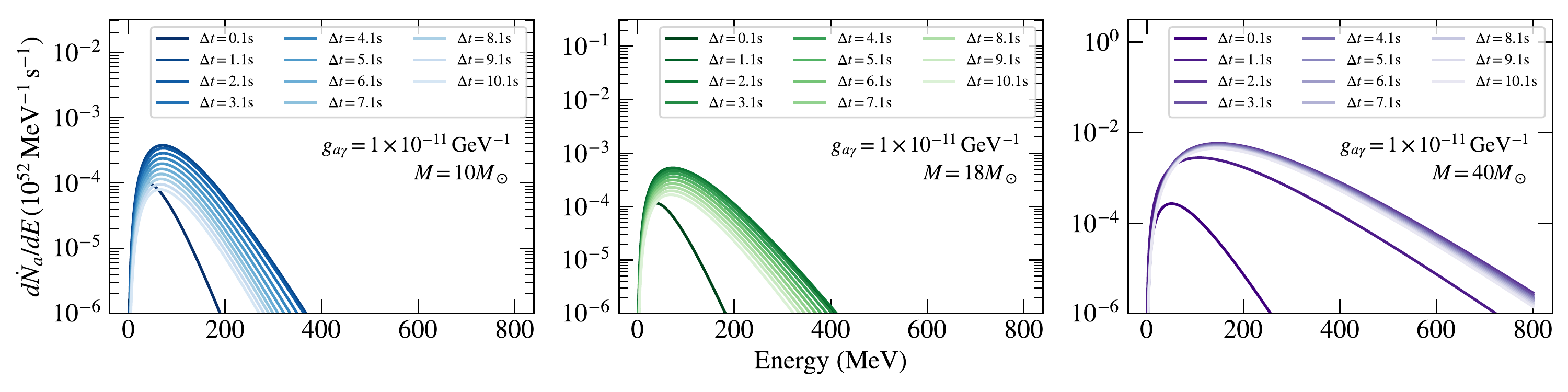}
    \caption{ALP production rate in a core collapse of stars with different progenitor masses at different times $\Delta t$ after the collapse.}
    \label{fig:alprate}
\end{figure*}

With these caveats in mind, the hotter and denser environment leads to an ALP production which is enhanced by more than an order of magnitude, see the theoretical spectra in Fig.~\ref{fig:alprate}. 
The effect on the limits is shown by the solid lines in Fig.~\ref{fig:sys-unc}.
While the limits improve only mildly when increasing all progenitor masses from 10 to $18\,M_\odot$, they improve by more than a factor of 2 when progenitors with masses of $40\,M_\odot$ are considered.
Changing the mass has the strongest effect on our bounds. 

Our constraints change to a lesser extend when a different model for the Galactic magnetic field or the the evolution of the optical emission are assumed. 
Concerning the former, we have repeated the analysis with the model presented in Ref.~\cite{2011ApJ...728...14P} (PTKN11) which in general predicts a lower conversion rate than the JF12 model due to the smaller $B$-field strength in the halo component and the absence of the ``X'' $B$-field component. 
Comparing the results for the two models (solid and dashed blue lines in Fig.~\ref{fig:sys-unc}) reveals, however, that the effect on the limits is limited and subdominant in comparison to the effect of the different progenitor masses and with respect to the spread in the limits depending on which SN is included (see Fig.~2 in the main part of the paper). 

Similarly, the change in the constraints is negligible when the exppow model is considered (dashed-dotted blue line in Fig.~\ref{fig:sys-unc}) to model the optical SN emission instead of the NiCo model.
The only difference between the NiCo and the exppow model for the \gr part of the analysis is that different GTIs are selected for the analysis. 
However, as already seen in Fig.~\ref{fig:all-gr-lc}, none of the GTIs shows evidence of a signal and hence the limits only change marginally.   
The same is true when we use the literature values for all SNe where available (dotted blue line in Fig.~\ref{fig:sys-unc}). 
We also report the values for $p_\mathrm{obs}$ and the highest found $\mathrm{TS}$ values found under different model assumptions in Tab.~\ref{tab:pobsTS}.
No significant detection of a signal is found in any model assumed. 
The overall probability to observe at least one SN is reported in the last row of Tab.~\ref{tab:pobsTS}. It is of the order of 90\,\% for all adopted models except when only literature values are used. 
Since there are fewer SNe in this sample for which the explosion time is estimated, the probability is decreased to 80\,\%.

\begin{turnpage}
\begin{table*}
\centering
\begin{scriptsize}
\begin{tabular}{l|cc|cc|cc|cc|cc|cc|cc}
\hline
\hline
\multirow{2}{*}{SN}
& \multicolumn{2}{|c}{$10\,M_\odot$, JF12, NiCo} 
& \multicolumn{2}{|c}{$10\,M_\odot$, JF12, SBO+NiCo} 
& \multicolumn{2}{|c}{$10\,M_\odot$, JF12, Exp. PL} 
& \multicolumn{2}{|c}{$18\,M_\odot$, JF12, NiCo} 
& \multicolumn{2}{|c}{$40\,M_\odot$, JF12, NiCo} 
& \multicolumn{2}{|c}{$10\,M_\odot$, PTKN11, NiCo} 
& \multicolumn{2}{|c}{$10\,M_\odot$, JF12, Lit. val.} 

\\
{} 
& $\mathrm{TS}_\mathrm{max}$ & $p_\mathrm{obs}$
& $\mathrm{TS}_\mathrm{max}$ & $p_\mathrm{obs}$
& $\mathrm{TS}_\mathrm{max}$ & $p_\mathrm{obs}$
& $\mathrm{TS}_\mathrm{max}$ & $p_\mathrm{obs}$
& $\mathrm{TS}_\mathrm{max}$ & $p_\mathrm{obs}$
& $\mathrm{TS}_\mathrm{max}$ & $p_\mathrm{obs}$
& $\mathrm{TS}_\mathrm{max}$ & $p_\mathrm{obs}$

\\
\hline
SN2009bb   & $2.59$ & $0.09$ & $4.48$ & $0.09$ & $4.48$ & $0.09$ & $2.53$ & $0.10$ & $1.16$ & $0.11$ & $2.59$ & $0.09$ & $0.07$ & $0.05$ \\
SN2009iz   & $0.00$ & $0.10$ & $0.00$ & $0.10$ & $7.84$ & $0.12$ & $0.00$ & $0.10$ & $0.00$ & $0.10$ & $0.00$ & $0.10$ & --- & $0.00$ \\
SN2009jf   & $7.03$ & $0.06$ & $7.03$ & $0.06$ & $5.59$ & $0.10$ & $7.29$ & $0.06$ & $6.30$ & $0.07$ & $7.03$ & $0.06$ & $7.03$ & $0.08$ \\
SN2010bh   & $0.00$ & $0.16$ & $0.00$ & $0.16$ & $0.00$ & $0.16$ & $0.00$ & $0.16$ & $0.00$ & $0.17$ & $0.00$ & $0.16$ & $2.68$ & $0.11$ \\
SN2010et   & $8.07$ & $0.10$ & $8.07$ & $0.10$ & $6.14$ & $0.09$ & $8.64$ & $0.10$ & $8.83$ & $0.12$ & $8.07$ & $0.10$ & $6.14$ & $0.09$ \\
SN2011bm   & $12.43$ & $0.12$ & $11.48$ & $0.12$ & $11.48$ & $0.12$ & $12.39$ & $0.12$ & $7.24$ & $0.13$ & $12.44$ & $0.12$ & $12.43$ & $0.07$ \\
SN2012P    & $5.31$ & $0.11$ & $5.31$ & $0.11$ & $4.35$ & $0.10$ & $5.26$ & $0.11$ & $4.00$ & $0.13$ & $5.31$ & $0.11$ & $3.76$ & $0.13$ \\
SN2012ap   & $1.30$ & $0.15$ & $1.64$ & $0.15$ & $1.43$ & $0.15$ & $1.09$ & $0.15$ & $0.80$ & $0.16$ & $1.30$ & $0.15$ & $1.43$ & $0.15$ \\
PTF12gzk   & $0.35$ & $0.13$ & $0.35$ & $0.14$ & $2.95$ & $0.13$ & $0.39$ & $0.13$ & $0.25$ & $0.14$ & $0.35$ & $0.13$ & $2.95$ & $0.10$ \\
iPTF13bvn  & $0.00$ & $0.09$ & $0.00$ & $0.09$ & $0.52$ & $0.07$ & $0.00$ & $0.09$ & $1.22$ & $0.11$ & $0.00$ & $0.09$ & $8.37$ & $0.05$ \\
SN2013ge   & $7.88$ & $0.12$ & $7.88$ & $0.12$ & $5.55$ & $0.13$ & $7.76$ & $0.12$ & $4.74$ & $0.14$ & $7.88$ & $0.12$ & $5.55$ & $0.14$ \\
SN2014L    & $3.01$ & $0.08$ & $3.01$ & $0.08$ & $3.01$ & $0.08$ & $3.08$ & $0.08$ & $2.74$ & $0.10$ & $3.01$ & $0.08$ & $8.23$ & $0.06$ \\
SN2014ad   & $2.78$ & $0.05$ & $5.74$ & $0.05$ & $5.84$ & $0.05$ & $2.98$ & $0.06$ & $3.74$ & $0.07$ & $2.77$ & $0.05$ & $5.84$ & $0.05$ \\
SN2015ap   & $2.67$ & $0.09$ & $2.67$ & $0.09$ & $4.14$ & $0.10$ & $2.89$ & $0.09$ & $5.53$ & $0.10$ & $2.67$ & $0.09$ & $4.14$ & $0.09$ \\
PTF15dtg   & $2.56$ & $0.10$ & $1.50$ & $0.13$ & $2.56$ & $0.12$ & $2.36$ & $0.10$ & $2.19$ & $0.11$ & $2.56$ & $0.10$ & $1.96$ & $0.11$ \\
SN2016bau  & $0.00$ & $0.04$ & $0.00$ & $0.05$ & $0.35$ & $0.08$ & $0.00$ & $0.04$ & $0.00$ & $0.04$ & $0.00$ & $0.04$ & --- & $0.00$ \\
SN2016blz  & $7.38$ & $0.14$ & $7.38$ & $0.14$ & $7.38$ & $0.14$ & $7.14$ & $0.14$ & $4.45$ & $0.15$ & $7.37$ & $0.14$ & --- & $0.00$ \\
SN2016coi  & $2.55$ & $0.11$ & $4.96$ & $0.11$ & $2.98$ & $0.11$ & $2.85$ & $0.11$ & $3.24$ & $0.13$ & $2.55$ & $0.11$ & $4.96$ & $0.11$ \\
SN2017ein  & $0.69$ & $0.15$ & $0.69$ & $0.14$ & $4.24$ & $0.15$ & $0.82$ & $0.15$ & $3.44$ & $0.16$ & $0.69$ & $0.15$ & $3.97$ & $0.11$ \\
SN2017fwm  & $0.14$ & $0.04$ & $0.14$ & $0.04$ & $0.24$ & $0.09$ & $0.16$ & $0.04$ & $0.77$ & $0.05$ & $0.14$ & $0.04$ & --- & $0.00$ \\
\hline
$P(N_\mathrm{SN} \geqslant 1)$ & \multicolumn{2}{|c}{0.89} 
& \multicolumn{2}{|c}{0.89} 
& \multicolumn{2}{|c}{0.90} 
& \multicolumn{2}{|c}{0.89} 
& \multicolumn{2}{|c}{0.91} 
& \multicolumn{2}{|c}{0.89} 
& \multicolumn{2}{|c}{0.79} 
\\
\hline
\end{tabular}

\caption{The maximum $\mathrm{TS}$ values found in the \gr analysis and probabilities that a particular SN has been observed with the LAT, $p_\mathrm{obs}$, for all SNe and tested model combinations. 
The last row reports the probability that at least one SN has been observed.
\label{tab:pobsTS}
}
\end{scriptsize}
\end{table*}
\end{turnpage}

\section{Sensitivity estimates for a growing SN sample}

In this section, we provide further details on how we estimate the sensitivity improvement of LAT observations with a growing sample of SNe. 

We artificially enlarge our SN sample by bootstrapping, i.e., we randomly draw $n$ SNe from our sample and add them to it, where $n=1,\ldots,80$. 
The results are shown in Fig.~\figsens.
The left panel shows how the median number of SNe that are included in the stacking of the individual limits improves with  growing sample size. 
Since the individual SNe have a probability of $\sim 10\,\%$ that they were in the field of view of the LAT during the core collapse, roughly 10\,\% of the total sample size are included in the limit stacking. 
Furthermore, the probability that at least one SN was in the field of view (orange line and right axis in the right panel of Fig.~\figsens) continuously improves and is above 99\,\% for a sample size $> 40$.

With a growing sample size and in the absence of ALPs, the constraints on the photon-ALP coupling will also improve. 
The level of improvement will depend on various factors such as the distance of the newly detected SNe and the exposure of the LAT during the most likely time of the core collapse. 
The limit improvement is shown in the left panel of Fig.~\figsens. 
If we bootstrap from the entire sample the limits improve between 10\,\% and 20\,\% for a sample size of 50.
On the other hand, if the newly detected SNe provide similar constraints to the ones obtained from iPTF13bvn, which provides the strongest limits in our sample, the constraints improve up to a factor of 2 for a sample size of 60. 
This is the most optimistic estimate given the constraints of the currently included SNe since we simply
added multiple copies of iPTF13bvn observations to the sample instead of bootstrapping.

To estimate the expected number of SNe Ib/c for ZTF at $z<0.02$, we use the software package \textsc{simsurvey}~\cite{2019JCAP...10..005F}.\footnote{\url{https://simsurvey.readthedocs.io/}}
This simulation tool takes into account the ZTF observing strategy, such as the cadence, filter choice, sky coverage and depth coverage of observations. 
To simulate the SNe Ib/c light curve model parameters,  e.g., peak magnitude, light curve width, or host galaxy extinction for each SN, they are drawn from a distribution modeling the SN Ib/c population. The SNe are distributed in redshift assuming a constant volumetric rate taken from the literature. \textsc{simsurvey} also predicts the phase at which a simulated SN was detected, where the phase is defined as the time of the second $5\,\sigma$ detection of the SN.  
This yields the results quoted in the main text of the paper. 

To estimate the number of SNe detected by the Rubin Observatory, we use the SN rate of Ref.~\cite{2019ApJS..243....6G},
which predicts a rate of $\sim138$ SNe Ib/c per year up to a redshift of 0.02 over the whole sky and $\sim60\,\mathrm{yr}^{-1}$ within the 18,000 square degree covered by the Wide, Fast, Deep (WFD) survey of the observatory. 
Given the Observatory's sensitivity, such close-by explosions could be detected minutes to hours after the shock breakout.
We further assume a Poisson distribution of SN events that are randomly distributed in time.
The time between revisits of observation fields (cadence time) is still highly debated and here
we set it to 3~days as proposed for the \texttt{minion\_1016} observation strategy~\cite{2017arXiv170804058L}, assuming that each SN will be detected regardless of the optical filter.
With these assumptions we find that the Rubin Observatory, on average, should detect close-by SNe Ib/c 1.5 days after the explosion and $\sim20$ SNe per year detected within 1~day after explosion. 

\begin{figure}
    \centering
    \includegraphics[width=.49\linewidth]{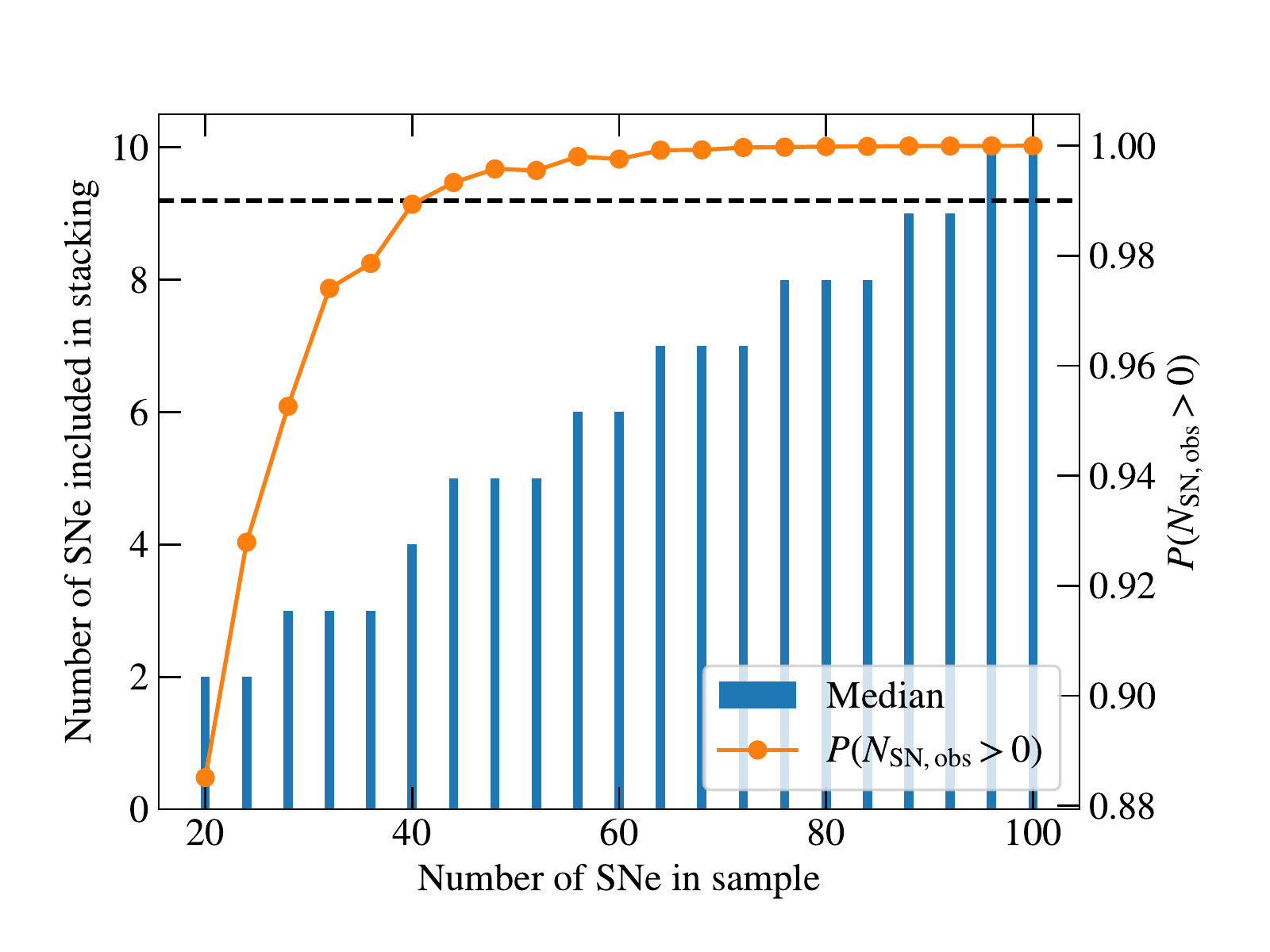}
    \includegraphics[width=.49\linewidth]{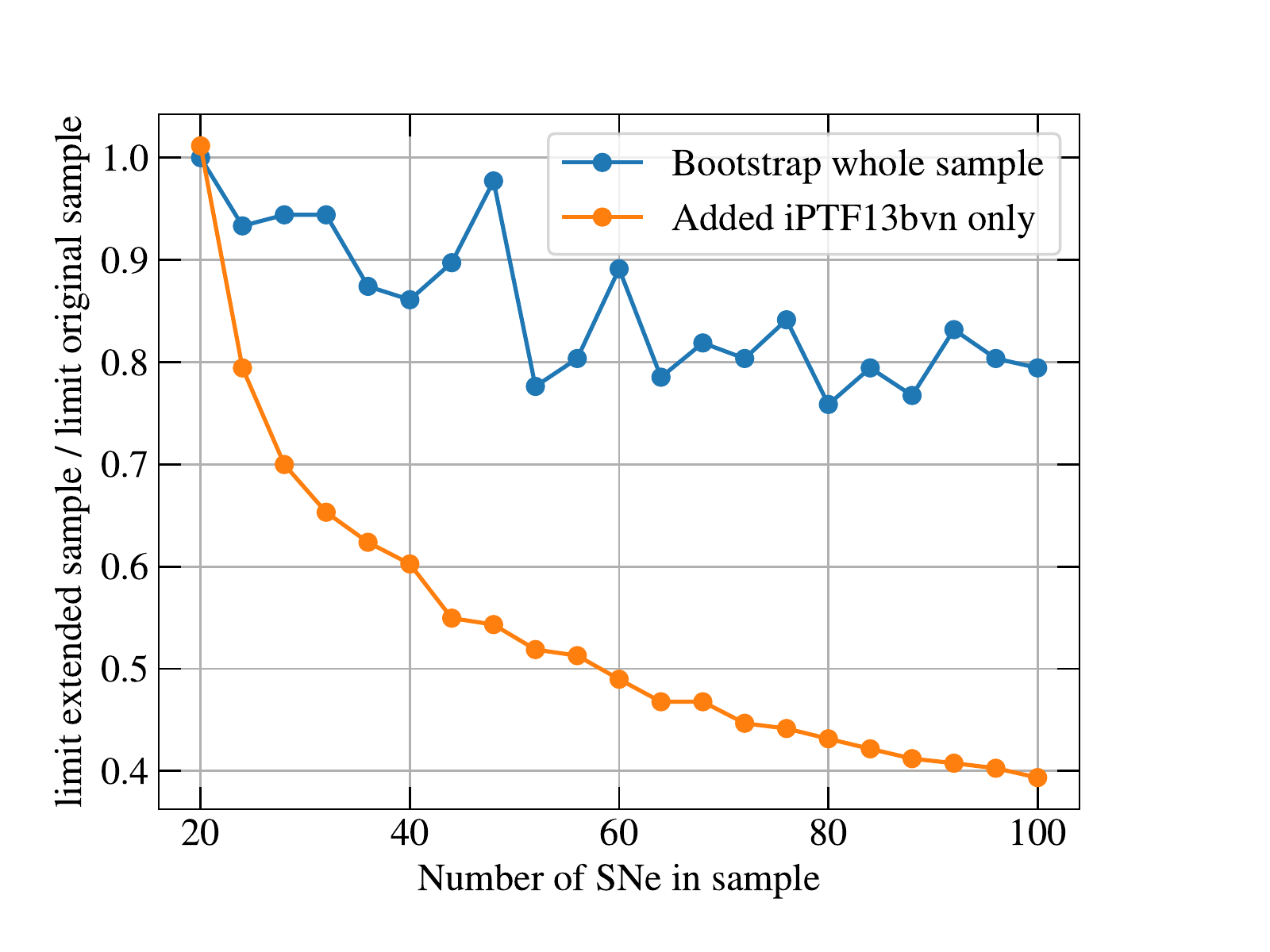}
    \caption{
    Estimates on sensitivity improvement with growing SN sample size. 
    \textit{Left:} The increasing median number of SNe included in the stacking (blue bars, left axis) as well as the probability that at least one SN was in the field of view of the LAT during the core collapse (orange line, right axis). 
    \textit{Right:} Anticipated fractional improvement of the constraints on ALPs if additional SNe are added to the sample by bootstrapping the entire sample (blue line) or by only adding iPTF13bvn (orange line) to the enlarged sample. 
    }
    \label{fig:sensitivity}
\end{figure}

We further note that core-collapse SNe detected in the field of view of the TESS satellite are of particular interest.
The satellite delivers light curves with a 30 minute cadence, which could therefore be used to determine the onset of the optical emission  with high accuracy. 
This has already been demonstrated for type~Ia SNe~\cite{2019arXiv190402171F}. 

\ifArxiv


\else

\bibliographystyle{JHEP}
\bibliography{references}
\end{document}

\fi

\fi

\end{document}
%